%% file: main.tex
\documentclass[aps,prb,twocolumn,nofootinbib]{revtex4-2}

\usepackage[a-1b]{pdfx}
\input{defs}

\hypersetup{
    colorlinks = true,
    linkbordercolor = {white}
}

\begin{document}
\title{Topological chiral superconductivity beyond pairing in a Fermi liquid 
}

\author{Minho Kim$^\dagger$}

\author{Abigail Timmel$^\dagger$}

\author{Long Ju}

\author{Xiao-Gang Wen$^\bigstar$}

\affiliation{Department of Physics, Massachusetts Institute of Technology,
Cambridge, Massachusetts 02139, USA\\
\\
$ ^\dagger $ \sf These authors contributed equally to this work.\\
$^\bigstar$ xgwen@mit.edu
}


\begin{abstract}

We investigate a mechanism to produce superconductivity by strong purely
repulsive interactions for flat dispersion $\veps \sim k^4$, without using
pairing instability in Fermi-liquid. The resulting superconductors break both
time-reversal and reflection symmetries in the orbital motion of electrons, and
exhibit non-trivial topological order. Our findings suggest that this
topological chiral superconductivity is more likely to emerge near or between
fully spin-valley polarized metallic phase and Wigner crystal phase.  These
topological chiral superconductors can be fully or partially spin-valley
polarized.  For  partial spin-valley polarization, the ratios of electron
densities associated with different spin-valley quantum numbers are quantized as
simple rational numbers.  Furthermore, many of these topological chiral
superconductors exhibit charge-4 or higher condensation, neutral quasiparticles
with fractional statistics, and/or gapless chiral edge states. Two of the
topological chiral superconductors are in the same phases as the
``spin''-triplet or spinless $p+ \textrm{i} p$ BCS superconductor, while others
are in different phases than any BCS superconductors. The same mechanism is also
used to produce anyon superconductivity between fractional anomalous quantum
Hall states in the presence of a periodic potential.

\end{abstract}

\maketitle

\setcounter{tocdepth}{1}
{\small \tableofcontents }

\section{Introduction}

After the discovery of superconductivity in 1911 \cite{O1122}, the standard BCS
mechanism for superconductivity was developed in 1957 \cite{BCS5775}, based on
the electron pairing instability of Fermi liquid, caused by an effective
attraction between electrons. In this paper, we explore a very different
mechanism of superconductivity, which is caused by strong purely repulsive
interaction.  The superconductivity from our mechanism is very different from
BCS superconductivity.

In fact, the mechanism based on the charged anyons in chiral spin liquid
\cite{KL8795,L8857,WWZ8913}, and related models
\cite{CWW8901,L9137,WZ9174,W9243,Lc9902287,KLW0902,TW1328}, belongs to this
class of mechanism (\ie driven by purely repulsive interactions).  In this
paper, we obtain superconductivity directly from electrons with repulsive
interaction, without going through charged anyons and the associated anyon
superconductivity.  The resulting superconductivity may not be associated with
electron pairing; charge-$4$ (and higher) condensation is also possible
\cite{KLW0902}.  As a result, the  resulting superconductivity usually carries
non-trivial topological order \cite{W8987,W9039}, which will be referred to as
topological chiral superconductivity.

The idea behind this non-BCS  mechanism is the following. We first assume that
electron hopping amplitude is complex, due to spontaneous time reversal
symmetry breaking and/or spin-orbit coupling.  We also assume that electron
interaction is larger than electron hopping energy.  In this case, when
electrons have an incommensurate density, they may not form a Fermi liquid.  

Certainly, when interaction is weak, electrons will form a Fermi liquid.
However, when interaction is strong, the electron motions are highly
correlated.\footnote{ We remark that correlated hopping was used to generate
effective attractions \cite{HM9049,GW180605990}, which may cause the pairing
instability of Fermi liquid.  In our work, correlated hopping directly leads to
the proposed chiral superconductivity without involvement of Fermi liquid and
its paring instability. } Since electron hopping is complex, the two electrons
exchange their place via correlated motion, and the phase factor can be
arbitrary.  In this case, electrons may forget their Fermi statistics.  Thus,
electrons may form a superconducting state via the mechanism of anyon or boson
superconductivity. The above idea is very rough, but may point to a right
direction \cite{TW1328}.  

In this paper, we discuss a concrete realization of the above idea in
2-dimensional space.  We argue that, for flat dispersion $\veps \sim k^4$,  a
strong repulsive interaction may cause chiral superconductivity that
spontaneously breaks time reversal and space reflection symmetry in orbital
motion of electrons. Other sources of time-reversal symmetry breaking and/or
spin-orbital coupling in orbital motion may further help this chiral
superconductivity.
Our theory is closely related to the theory of anyon superconductivity developed in
\Rf{WZ9174,TW1328}. Since a fermion is a special case of anyon, the theory of
anyon superconductivity applies to fermion superconductivity with little change. 


Strong repulsive Coulomb interaction favor
fully spin-valley polarized Fermi liquid
(referred to as quarter Fermi liquid) and Wigner crystals, which have both been observed in
experiments at low densities where the interaction effect is
strong\cite{HJ240815233}.  Furthermore, a strong coupling superconducting state
that \emph{breaks time reversal and reflection symmetries} in electron orbital
motion was observed in \Rf{HJ240815233} between quarter Fermi liquid and Wigner
crystal, in tetralayer rhombohedral-stacked graphene without Moire pattern.
Other superconductivities were also observed in bilayer
\cite{CJ180302342,ZN220505087,HY230300742, LL240504479}, trilayer
\cite{ZY210607640,PY240810190}, and Moire-tetralayer \cite{CY240812584}
rhombohedral-stacked graphene systems. 

Some BCS-pairing mechanisms for those
superconducting states were explored in
\Rf{DA240617036,CD220609922,JG221002915,WP230200682,DL230409812,DL220513353,CL240419237,CS240906701,GF240913829}.
In this paper, we take a very different approach of using only strong repulsive
interaction.  We find that, for flat dispersion $\veps \sim k^4$, a repulsive
Coulomb interaction may leads to chiral superconductors, which are different
from BCS superconductors since they are not induced by pairing instability of
Fermi liquid. 

Furthermore, based on our numerical calculations, we find that, for $\veps \sim k^2$ dispersion, the
proposed chiral superconductors exhibit lower energy than the quarter Fermi
liquid only at very low densities. However, at these low densities, we expect
the Wigner crystal phase to have an even lower energy. In contrast, for $\veps \sim k^4$
dispersion, the chiral superconductors demonstrate lower energy than the
quarter Fermi liquid at densities near the transition to the Wigner crystal
phase. This observation highlights that the $\veps \sim k^4$ dispersion is a crucial factor
in realizing chiral superconductors driven by Coulomb interactions.

Some of chiral superconductors are fully spin-valley polarized or half
spin-valley polarized (with two spin-valley components present and at equal
density), while others are spin-valley un-polarized (with all four spin-valley
components at equal density).  The fully spin-valley polarized chiral
superconductors are in the same phases as the spinless $p+\ii p$ BCS
superconductor \cite{RG0067}.  For partial spin-valley polarization, the ratio of different
species of electrons are quantized as simple rational numbers. In this case, in
contrast to BCS superconductors of quarter Fermi liquid, the transition from
spin-valley partially-polarized chiral superconductors to quarter Fermi liquid
cannot be continuous at zero temperature in the clean limit.  

Also, as we lower the electron density, a topological chiral superconductor is
likely to change into a Wigner crystal, via a first order transition.  Thus
topological chiral superconductivity is more likely to appear near the
transition between quarter Fermi liquid and Wigner crystal, since all those
phases are driven by strong repulsive interactions. As a strong coupling
superconductor, the coherence length of a chiral superconductor is about the
same as the inter-electron separation.

All these chiral superconductors are topological (\ie carry non-trivial
topological order, where another example was given in \Rf{SF0050}). As a result,
they carry neutral excitations with fractional self/mutual statistics and/or
gapless chiral edge modes. They all break time reversal and space reflection
symmetry. We expect those symmetry breaking and the associated chiral
correlation (such as chiral edge state) to persist even above the
superconducting transition temperature $T_c$. Just above $T_c$, the chiral
superconductors should have non-zero Hall conductance of order $e^2/h$. Except
two chiral superconductors, other chiral superconductors are not in the same
phases as any BCS superconductors. For example, many of chiral superconductors
have charge-4 \cite{KLW0902} or higher condensation. 

Our calculation is not reliable enough to predict if a chiral superconductor
can appear or not (\ie can have energy below both quarter Fermi liquid and
Wigner crystal). However, if a chiral superconductor (\ie a superconductor that
breaks time reversal and reflection symmetry) is observed in experiments near
quarter Fermi liquid and Wigner crystal as in \Rf{HJ240815233}, our calculation
suggests that it may be a topological chiral superconductor, with properties
described above.  

The calculated ground-state energies of these topological chiral superconductors
are lower, but still close to Hartree-Fock energy of a quarter Fermi liquid. To
achieve an even lower energy, additional time-reversal symmetry breaking and/or
spin-orbit interaction may be helpful.  The strong geometric phase curvature
near the bottom of the graphene band also contributes to breaking time-reversal
symmetry through partial spin-valley polarization.

We remark that, although most topological chiral superconductors belong to
different phases than BCS superconductors, two of the topological chiral
superconductors (which we denote as the $K_{2a}$-chiral superconductor and
Pfaffian chiral superconductor) belong to the same phase as the ``spin''-triplet
or spinless $p+\ii p$ BCS superconductor. Here ``spin'' corresponds to a pair of
spin-valley quantum numbers, and may not be the electron spin. Such
$K_{2a}$-chiral superconductor and Pfaffian chiral superconductor can be induced
by a purely repulsive interaction. They can also be induced by a pairing
instability of a half Fermi liquid (for $K_{2a}$-chiral superconductor) or
quarter Fermi liquid (for Pfaffian chiral superconductor) caused by an effective
attractive interaction. Here, half Fermi liquid refers to a Fermi liquid formed
by two species of electrons of equal density. We note that very recently, a
superconducting state has been observed next to a half Fermi liquid in
rhombohedral trilayer graphene in \Rf{PY240810190}. The $K_{2a}$-chiral
superconductor and Pfaffian chiral superconductor also have similarly low
energies as other chiral superconductors at lower densities. Thus, it may also
appear near the quarter Fermi liquid and Wigner crystal, as in rhombohedral
tetralayer graphene in \Rf{HJ240815233}. 

In the second part of paper, we will use the same method to discuss possible
anyon superconductivity between two fractional quantum anomalous Hall (FQAH)
states. The periodic potential in FQAH states play a crucial role for the
appearance of anyon superconductivity.

\section{Chiral superconductivity driven by purely repulsive interactions}

To construct a topological chiral superconducting state, let us consider a simple case of
spin-$\frac12$ electrons in 2-dimension space.  We view each electron as a
bound state of a boson and $2\pi$-flux \cite{ZHK8982}.  We then smear the
$2\pi$-flux into a uniform ``magnetic'' field $\tl b$.  In this case,
the interacting spin-$\frac12$ electrons are effectively described by
interacting spin-$\frac12$ bosons in a uniform magnetic field, with a filling
fraction $\nu=1$.  The interacting bosons can form various states that
correspond to various states of interacting electrons.

When repulsive interaction is strong, the interacting bosons may form an
incompressible fractional quantum Hall (FQH) state.  In this case, the only low
energy fluctuations are the co-fluctuations of the boson density and ``magnetic''
field $\tl b$ keeping the filling fraction $\nu=1$ fixed.  Such
co-fluctuations are gapless and are the only gapless modes of the system. In
this case, the  system is in a superconducting state (\ie a superfluid state)
\cite{WZ9040}.

We remark that this FQH state does not emerge from single-particle Landau levels produced by an
external magnetic field; it has purely many-body origins.  Under broken
time-reversal symmetry, attaching $2\pi$ flux to the fields is an allowed operation which may 
lower the energy of the state.  We model this with FQH wavefunctions, which are the simplest 
way to implement flux attachment while keeping density uniform.
Indeed, the single-particle orbitals are not eigenstates of a kinetic energy operator without an
external magnetic field, but they enforce 1. statistics consistent with flux attachment, 2. zeros in
the wavefunction that favor repulsive interactions, and 3. uniform density apart from the superfluid
mode controlled by a length scale $l_b$.  The precise form of the many-body wavefunction may differ 
from the Laughlin states studied here, but we use these as representatives for phases which may arise 
from strong correlations.

For interacting spin-$\frac12$ bosons with filling fraction $\nu=1$,
the most natural FQH state is given by the following wave function
\begin{align}
 \Phi(z_i^\up,z_i^\down)
= 
\ee^{-\frac{ \sum_i |z_i^\up|^2 +|z_i^\down|^2}{4l_b^2}}
\prod_{i<j} (z_i^\up - z_j^\up)^2  (z_i^\down - z_j^\down)^2
\end{align}
where $z_i^\up,z_i^\down$ are complex numbers describing the boson coordinates,
and $l_b$ is the length scale of electron separation.  Such a bosonic FQH state
corresponds to the following electron wave function 
\begin{align}
\label{Psi12}
 \Psi(z_i^\up,z_i^\down)
&= 
\ee^{-\frac{ \sum_i |z_i^\up|^2 +|z_i^\down|^2}{4l_b^2}}
\prod_{i<j} (z_i^\up - z_j^\up)^2  (z_i^\down - z_j^\down)^2
\nonumber\\
&
\prod_{i<j} (z_i^{\up *} - z_j^{\up *})  (z_i^{\down *} - z_j^{\down *})
\prod_{i,j} (z_i^{\up *} - z_j^{\down *}),  
\end{align}
where the factor $\prod_{i<j} (z_i^{\up *} - z_j^{\up *})  (z_i^{\down *} -
z_j^{\down *}) \prod_{i,j} (z_i^{\up *} - z_j^{\down *})$ is the wave-function
representation of the flux smearing operation.

The superconducting mode in this system is a co-fluctuation of up and down spins.
Increasing the density of up spins in a region increases the effective magnetic 
field seen by the down spins, causing their density to increase to keep the
filling fraction fixed.  The increased down spin density in turn increases the 
effective magnetic field seen by the up spins, preserving their original filling fraction.
These density fluctuations appear as fluctuations of $l_b$ in the wavefunction.

We now examine the energetics of this state. Let $N_\up$ ($N_\down$) be the total 
number of spin-$\up$ (spin-$\down$)  electrons. The total angular momentum of the above state is
\begin{align}
 L_z^\text{tot} &= 
 N_\up (N_\up-1) 
+N_\down (N_\down-1) 
\nonumber\\
&\ \ \ \
- \frac12 N_\up (N_\up-1) 
- \frac12 N_\down (N_\down-1) -N_\up N_\down
\nonumber\\
&= \frac12 (N_\up-N_\down)^2
-\frac{N_\up+N_\down}{2}.
\end{align}
We see that each pair of spin-$\frac12$ electrons has an angular momentum
$L_z=-1$, just like a spin triplet $p$-wave paired superconducting state.  It is crucial
that the leading $N$-contribution to the angular momentum of each electron is
cancelled completely.  Otherwise, the
kinetic energy of the wave-function $\Psi(z_i^\up,z_i^\down) $ will be too high. 

This cancellation can be seen by comparing the total angular momenta
for $N_\up=N_\down$ system and $N_\up=N_\down+1$ system.  We can also fix the position of all other electrons and
consider the motion of, say, first spin-$\up$ electron. From the wave function, we see that other spin-$\up$ electrons
behave like $2\pi$-flux quanta and other spin-$\down$ electrons
behave like $-2\pi$-flux quanta. The first spin-$\up$ electron sees a zero average ``magnetic'' field. Thus its angular 
momentum does not contain a linear-$N$ term.

The electron wave-function $\Psi(z_i^\up,z_i^\down)$ has a third order zero
between two spin-$\up$ (spin-$\down$) electrons, and has a first order zero
between a spin-$\up$ and a spin-$\down$ electrons.  So
$\Psi(z_i^\up,z_i^\down)$ has a reduced interaction energy compared to Fermi
liquid state.  To estimate this reduction very roughly,
we assume the interaction energy per electron to be
\begin{align}
 E_\text{int} = U \left\<\frac{1}{n+1}\right\>
\end{align}
where $U$ is the interaction strength at distance $l_b$ and $\<\frac{1}{n+1}\>$
is the average of inverse order of zeros (shifted by 1).  For example, the
wave-function $\Psi(z_i^\up,z_i^\down)$ has an interaction energy per electron
\begin{align}
\label{Eint}
 E_\text{int} = U \frac{\frac1{3+1}+\frac1{1+1}}{2} = \frac 38 U.
\end{align}
In comparison, a Fermi liquid of spin-$\frac12$ electrons
 has an interaction energy per electron
\begin{align}
 E_\text{int} = U \frac{\frac1{1+1}+\frac1{0+1}}{2} = \frac 34 U,
\end{align}
since the order of zeros between two spin-$\up$ (spin-$\down$) electrons is 1
and the order of zeros between a spin-$\up$ and a spin-$\down$ electrons is 0,
for the Fermi liquid. 

The electron wave-function $\Psi(z_i^\up,z_i^\down)$ has a higher kinetic energy
compared to the Fermi liquid, since each electron has a larger momentum.  The
typical momentum of each electron can be estimated as
\begin{align}
 p = (\< n\> n_e)^{1/2}, 
\end{align}
where $\< n\>$ is the average of order of zeros and $n_e$ is the electron density.
For $\Psi(z_i^\up,z_i^\down)$, we have
\begin{align}
\label{kterm}
 p =  (n_e (3+1)/2)^{1/2} = \sqrt{2} n_e^{1/2},
\end{align}
while for Fermi liquid, we have
\begin{align}
 p =  n_e^{1/2}((1+0)/2)^{1/2} = \frac{1}{\sqrt{2}} n_e^{1/2},
\end{align}

If the electron has an effective mass $m$, the chiral superconducting state will have
an energy (per electron)
\begin{align}
\label{Esuper}
 E_\text{FQH} =
\frac{n_e}{m} +  \frac 38 U ,
\end{align}
while the Fermi liquid 
will have
an energy (per electron)
\begin{align}
 E_\text{Fermi} =
\frac{n_e}{4m} + \frac 34 U . 
\end{align}
This will give us some idea when the chiral superconducting state is favored.
Certainly, the above calculation is crude, but we present it here to
illustrate the reasoning behind our idea.  A more rigorous calculation
is included in the appendix.

Our above construction of electron chiral superconducting states also applies to the
situation where we have several species of electrons labeled by $I$. In the above case,
$I=\up,\down$.  We can more generally view an electron as a
bound state of a boson and $2\pi k_f$-flux for odd $k_f$, or a bound state of a
fermion and $2\pi k_f$-flux for even $k_f$.
The chiral superconducting state, from the flux smearing, is given by
\begin{align}
\label{PsiK}
 \Psi(z_i^I)
&= 
\ee^{-\frac{ \sum_{i,I} |z_i^I|^2}{4l_b^2}}
\prod_{i<j,I} (z_i^I - z_j^I)^{K^\text{QH}_{II}}  
\prod_{i,j,I<J} (z_i^I - z_j^J)^{K^\text{QH}_{IJ}}  
\nonumber\\
&\ \ \ \
\prod_{i<j,I} (z_i^{I*} - z_j^{I*})^{k_f} 
\prod_{i,j,I<J} (z_i^{I*} - z_j^{J*})^{k_f}
\end{align}
The exponents, $K^\text{QH}_{IJ}$ form a $K$-matrix, which is a symmetric integral
matrix, which describes a filling fraction $\nu=\frac1{k_f}$ FQH state
\cite{WZ9290}.  Thus $K$ must satisfy
\begin{align}
\label{Keq3}
& q_I (K^{\text{QH}})^{-1}_{IJ} q_J = \frac 1{k_f}, \ \ \ \ q_I = 1,
\nonumber\\
& f_I = k_f (K^{\text{QH}})^{-1}_{IJ} q_J \geq 0.
\end{align}
$f_I$ is the fraction of species-$I$ electrons, and so
$f_I\geq 0$.  Also the diagonal elements of the $K$-matrix are even for odd
$k_f$, to describe a FQH state of bosons.  The diagonal elements of the
$K$-matrix contain odd integers for even $k_f$ to describe a FQH state of
fermions. $K^\text{QH}_{IJ}$ can also be negative, in which case
$(z_i^I - z_j^J)^{K^\text{QH}_{IJ}}$ is understood as $(z_i^{I *} - z_j^{J
*})^{-K^\text{QH}_{IJ}}$.

We can more clearly distinguish the exponents of the holomorphic and antiholomorphic
components of the wavefunction by defining
\begin{align}
 K^+_{IJ} &=
\begin{cases}
 K^\text{QH}_{IJ} &\text{ if } K^\text{QH}_{IJ}>0\\
 0 &\text{ if } K^\text{QH}_{IJ}<0\\
\end{cases}
,
\nonumber\\
 K^-_{IJ} &=
\begin{cases}
 k_f &\text{ if } K^\text{QH}_{IJ}>0\\
 k_f-K^\text{QH}_{IJ} &\text{ if } K^\text{QH}_{IJ}<0\\
\end{cases}
\end{align}
where $K^+$ and $K^-$ are non-negative integral matrices. The wave-function $\Psi$ in \eqref{PsiK} becomes
\begin{align}
\label{PsiKpm}
 \Psi(z_i^I)
&=
\ee^{-\frac{ \sum_{i,I} |z_i^I|^2}{4l_b^2}}
\prod_{i<j,I} (z_i^I - z_j^I)^{K^+_{II}}  
\prod_{i,j,I<J} (z_i^I - z_j^J)^{K^+_{IJ}}  
\nonumber\\
&\ \ \ \ 
\prod_{i<j,I} (z_i^{I *} - z_j^{I *})^{K^-_{II}}  
\prod_{i,j,I<J} (z_i^{I *} - z_j^{J *})^{K^-_{IJ}} , 
\end{align}

When both $K^+_{IJ}$ and $K^-_{IJ}$ are nonzero for a pair $IJ$, the
wave-function $\Psi$ contains a factor $|z_i^I - z_j^J|^2$. We can modify this
factor to $|z_i^I - z_j^J|^{2\al}$, and deform $\al$, trying to lower the energy
of the chiral superconductor further. Since there is no phase winding protecting
these zeros, we expect such a deformation to be a smooth deformation, that does
not change the phase of the ground state. 

We find that, at low densities, the ground state energy can be lowered if $K^+_{IJ}$ and $K^-_{IJ}$
are increased (or decreased) to sum to the maximum value of $|K^+_{IJ} - K^-_{IJ}|$,
which is referred to as $K_\text{max}$.
Thus, we will consider the following many-body wave-function
for our chiral superconductors characterized by a symmetric integral $K$-matrix
of odd diagonals:
\begin{align}
\label{PsibarK}
 \Psi(z_i^I)
&=
\ee^{- \sum_{i,I} \frac{|z_i^I|^2}{4l_I^2}}
\prod_{i<j,I} (z_i^I - z_j^I)^{\bar K^+_{II}}  
\prod_{i,j,I<J} (z_i^I - z_j^J)^{\bar K^+_{IJ}}  
\nonumber\\
&\ \ \ \ 
\prod_{i<j,I} (z_i^{I *} - z_j^{I *})^{\bar K^-_{II}}  
\prod_{i,j,I<J} (z_i^{I *} - z_j^{J *})^{\bar K^-_{IJ}} , 
\end{align}
where
\begin{align}
     &\bar K^+_{IJ} \geq 0,\ \ 
     \bar K^-_{IJ} \geq 0, \ \ K = K^+ - K^- = \bar K^+ - \bar K^-, 
     \nonumber\\
     &\bar K^+_{IJ} + \bar K^-_{IJ} = K_\text{max}
     .
\end{align}
Also, note that each species of electrons can have its own ``magnetic length'' $l_I$, which can be fully determined by
by $\bar K^\pm$ (see Appendix).
For
example, after removing the ``unnecessary'' zeros, the wave
function \eqref{Psi12} is simplified to 
\begin{align}
\label{PsiK2a}
 \Psi(z_i^\up,z_i^\down)
&= 
\ee^{-\frac{ \sum_i |z_i^\up|^2 +|z_i^\down|^2}{4l_b^2}}
\prod_{i<j} (z_i^\up - z_j^\up)  (z_i^\down - z_j^\down)
\nonumber\\
&
\prod_{i,j} (z_i^{\up *} - z_j^{\down *}).  
\end{align}

We will use \eqref{PsibarK} as a trial wave function for the associated chiral superconductor. Such a wave function is determined by $\bar K^\pm$, which must
satisfy some conditions, as discussed in detail in the Appendix. In the main text, we will just summarize the results. First, 
\begin{align}
 K_{IJ} = \bar K^+_{IJ} -\bar K^-_{IJ}
\end{align}
must be a symmetric integer matrix with odd diagonal elements, so that the wave function
is single-valued and anti-symmetric.  

The wave function \eqref{PsibarK} has total angular momentum
\begin{align}
  & \ \ \ \
  \sum_{I} \frac{N_I(N_I-1)}{2}   K_{II}   + \sum_{I<J} N_I N_J K_{IJ}  
  \nonumber \\
  & = \frac12 \sum_{I,J} N_I  K_{IJ} N_J - \frac12 \sum_{I} N_I  K_{II} 
  \nonumber \\
  & = \frac {N^2}2 \sum_{I,J} f_I  K_{IJ} f_J - \frac N2 \sum_{I} f_I  K_{II} 
\end{align}
where $N_I$ is the number of species-$I$ electrons, $N=\sum_I N_I$, and $f_I =
\frac{N_I}N$.  We remark that those $N_I$ electrons occupy a circular area with
an area $A_I$. We adjust $l_I$ in the wave function such that $A_I$ for
different species are all equal $A_I=A$.

In order to describe a
superconducting or superfluid state, $K$ is required to have a single zero
eigenvalue so that the total angular momentum does not contain the $N^2$ term:
\begin{align}
\label{Kf}
\sum_J K_{IJ} f_J = 0
\end{align}
The corresponding eigenvector $f_I$ describes a gapless mode within the otherwise
gapped chiral superconducting state.
In order to be a superconducting state, we also require the electron density fluctuations giving
rise to this mode to be net positive; otherwise, it would describe a charge-neutral
superfluid mode instead of a superconductor. 
$f_I$ is the fraction of species-$I$ electrons.
Thus, we also require
\begin{align}
\label{fpos}
 f_J = \text{all positive},
\end{align}
By definition
\begin{align}
 \sum_I f_I = 1.
\end{align}
We also note that the average angular momentum per electron is given by
\begin{align}
  \<L\> = - \frac12 \sum_{I} f_I  K_{II} 
\end{align}
which is a topological invariant of the chiral superconductor.

There are many $K$-matrices that satisfy \eqref{Kf} and
\eqref{fpos}.  To determine which are favorable, we first quantify the ground state energy 
of the chiral superconductor \eqref{PsibarK} more precisely.
As described in the introduction, we consider an electron dispersion $\varepsilon = 
c_2 k^2 + c_4 k^4 $, where the quartic term further flattens the bottom of the band compared
to a purely quadratic dispersion.  From a derivation carried out in the Appendix, 
it can be shown that the kinetic energy per electron has the form (\ref{KE})
\begin{align}
E_\text{kin} &=  
 2\pi n_e c_2 Z_2
+(2\pi n_e)^2 c_4 Z_4,
\end{align}
where $n_e$ is the total density of the electrons, and $Z_2,\ Z_4$ are dimensionless parameters
with the form (\ref{Zs})
\begin{align}
\label{Z2Z4}
 Z_2 & = \sum_{IJ}  f_I \bar K_{IJ} f_J + \sum_I f_I \frac{2g_{2, I}}{\pi n_e},
\nonumber\\
 Z_4 & = \sum_I 2 f_I(\sum_J f_J \bar K_{IJ})^2 + \sum_I f_I \frac{16g_{4, I}}{\pi^2 n_e^2},
\nonumber\\
 \bar K & \equiv \bar K^+_{IJ} +\bar K^-_{IJ}.
\end{align}
The dimensionless ratios, $\frac{g_2}{n_e}$ and $\frac{g_4}{n_e^2}$, are
computed numerically via Monte Carlo method, with a current error about 10\% 
(see Appendix).

If we assume the electrons interact via Coulomb interaction $\frac{e^2}{\eps r}$, then the interaction energy per electron is given by (\ref{eq:EandV})
\begin{align}
 E_\text{int} 
&=
\frac{e^2\sqrt{n_e}}{\eps} V, \ \ \
V\equiv 
\sum_{IJ} f_I f_J V_{IJ}  
\end{align}
where
\begin{align}
 V_{IJ} \equiv \int
\dd^2 z \ \frac{ \sqrt{n_e}}{ 2|z|} (g_{IJ}(z)-1) .
\end{align}
and $g_{IJ}(z)$ is the electron pair distribution function which must
be computed numerically.  We find the following approximate fitting for $V_{IJ}$ (\ref{V_approx})
\begin{align}
V_{IJ} =
\begin{cases}
 \frac{1}{\sqrt{f_I}}(-1.830 + \frac{0.408 }{\bar K_{II} + 0.433}),& I=J\\
 \Theta(\bar K_{IJ}) (-1.093 + \frac{0.117 }{\bar K_{IJ} - 0.596}), &  I\neq J\\
\end{cases}
\end{align}
with error $\sim 0.03$ (see Appendix), where $\Theta(0) = 0$ and $\Theta(x>0)
=1$. When $\bar K_{IJ}=K_\text{max}$, the energy $E_\text{int}$ happen to be the
same as the one component case $\bar K = (K_\text{max})$. We will use this
property to compute $E_\text{int}$ for the case $\bar K_{IJ}=K_\text{max}$.

To determine which
$K$-matrices give rise to the most stable chiral superconductors, 
we compute $Z_2$, $Z_4$, and $V$.
The total energy per electron is given by
\begin{align}
\label{EtotK}
 E_\text{tot}^K(n_e) =  
\frac{e^2}{\eps} n_e^{1/2} V + c_\ga (2\pi n_e)^{\ga/2} Z_\ga
,
\end{align}
where we have assumed the kinetic energy of an electron to be $c_\ga k^\ga$,
$\ga =2,4$.

For two species of electrons, we have
\begin{align}
 \label{K2a}
K_{2a} &=
\begin{pmatrix}
 1&  -1\\
 -1&  1\\
\end{pmatrix} , \ \ \
f_I = \left(\tfrac12,\tfrac12\right),
\\
V & = -1.5705, \ \ \
Z_2  = 1+2.19, \ \ \ 
Z_4  = 2+1.12. 
\nonumber
\end{align}
\begin{align}
 \label{K2b}
K_{2b} &=
\begin{pmatrix}
 3&  -3\\
 -3&  3\\
\end{pmatrix} , \ \ \
f_I = \left(\tfrac12,\tfrac12\right),
\\
V & = -1.7116, \ \ \
Z_2  = 3+20.80, \ \ \ 
Z_4  = 18+10.03. 
\nonumber
\end{align}
\begin{align}
 \label{K2c}
K_{2c} &=
\begin{pmatrix}
 5&  -5\\
 -5&  5\\
\end{pmatrix} , \ \ \
f_I = \left(\tfrac12,\tfrac12\right),
\\
V & = -1.7553, \ \ \
Z_2  = 5+52.50, \ \ \ 
Z_4  = 50+24.87. 
\nonumber
\end{align}
For three species
\begin{align}
 \label{K3a}
K_{3a} &=
\begin{pmatrix}
-3 & 2 & -1 \\ 
2 & -1 & 0 \\ 
-1 & 0 & 1 \\ 
\end{pmatrix} , \ \
f_I = \left(\tfrac14, \tfrac12, \tfrac14\right),
\\
V & =  -1.7116, \ \ \
Z_2  = 3+3.37, \ \ \ 
Z_4  =18+1.69. 
\nonumber
\end{align}
\begin{align}
 \label{K3b}
K_{3b} &=
\begin{pmatrix}
-3 & 1 & 0 \\ 
1 & 1 & -2 \\ 
0 & -2 & 3 \\ 
\end{pmatrix} , \ \
f_I = \left( \tfrac16, \tfrac12, \tfrac13 \right),
\\
V & = -1.7116, \ \ \ 
Z_2  =3+ 5.71, 
\ \ \
Z_4  =18+2.78. 
\nonumber
\end{align}
\begin{align}
 \label{K3c}
K_{3c} &=
\begin{pmatrix}
-3 & -1 & 5 \\ 
-1 & 3 & -5 \\ 
5 & -5 & 5 \\ 
\end{pmatrix} , \ \
f_I = \left( \tfrac14, \tfrac12, \tfrac14 \right),
\\
V & = -1.7553, \ \ \ 
Z_2  =5+ 31.32, 
\ \ \
Z_4  =50+14.82. 
\nonumber
\end{align}
For four species of electrons,
\begin{align}
 \label{K4a}
K_{4a} &=
\begin{pmatrix}
-1 & 1 & -2 & 2 \\ 
1 & -1 & 3 & -3 \\ 
-2 & 3 & 1 & -2 \\ 
2 & -3 & -2 & 3 \\
\end{pmatrix} , \ \
f_I = \left(\tfrac14,\tfrac14,\tfrac14,\tfrac14\right),
\\
V &  = -1.7116, \ \ \
Z_2  =3+9.82, \ \ \ 
Z_4  =18+4.65 . 
\nonumber
\end{align}
\begin{align}
 \label{K4b}
K_{4b} &=
\begin{pmatrix}
-1 & -1 & -1 & 3 \\ 
-1 & 1 & 3 & -3 \\ 
-1 & 3 & 1 & -3 \\ 
3 & -3 & -3 & 3 \\
\end{pmatrix} , \ \
f_I = \left(\tfrac14,\tfrac14,\tfrac14,\tfrac14\right),
\\
V &  = -1.7116, \ \ \
Z_2  =3+12.31, \ \ \ 
Z_4  =18+5.75 , 
\nonumber
\end{align}
\begin{align}
 \label{K4c}
K_{4c} &=
\begin{pmatrix}
-1 & -1 & 1 & 1 \\ 
-1 & -1 & 2 & 0 \\ 
1 & 2 & -1 & -2 \\ 
1 & 0 & -2 & 1 \\ 
\end{pmatrix} , \ \
f_I = \left(\tfrac14,\tfrac14,\tfrac14,\tfrac14\right),
\\
V &  = -1.6628, \ \ \
Z_2  =2+3.01, \ \ \ 
Z_4  =8+1.45. 
\nonumber
\end{align}
In the above, the $\bar K$ and $g_{2,4}$
contributions to $Z_{2,4}$ was separated by the sum.

Let us now consider tetralayer rhombohedral graphene to compare the above
states with competing Fermi liquid states.  The electron density $n_e$,
measured from charge neutrality, is of order $ 10^{12} \text{cm}^{-2}$.  We
will model the electron interaction by a screened Coulomb interaction. Usually,
the electron dispersion has a form $\varepsilon = c_2 k^2 + c_4 k^4 \cdots$.
We may use a displacement field to fine tune the dispersion to make $c_2=0$. So
$\ga$ in \eqref{EtotK} can be chosen to be $\ga=4$. We will also consider
$\ga=2$ case.

There are 4 species of electrons, carrying a spin $\al =\up,\down$
index and a valley $a =1,2$ index.  Those electrons may form a so called
``full'' Fermi liquid, in weak interaction limit, where all four species of
electrons have the same density.  But we are interested in the limit of strong
repulsive interaction. In this case, as indicated by experiments
\cite{HJ230503151}, the electrons form a so called ``quarter'' Fermi liquid,
where only 1 species of electrons is present, or a so called ``half'' Fermi
liquid, where only 2 species of electrons are present.

The Hartree-Fock energy for the experimentally observed quarter Fermi liquid 
is given by \eqref{EtotK} with (see eqn. (13) in \cite{TC8905})
\begin{align}
 V^\text{QFL} &= -\frac{8}{3 \sqrt{\pi} } 
 = -1.5045, 
\nonumber\\
Z_2^\text{QFL}  &= 
\frac{\int_0^1 \frac{2\pi k\dd k}{(2\pi)^2}\ k^2}
{(2\pi \int_0^1 \frac{2\pi k\dd k}{(2\pi)^2})  \int_0^1 \frac{2\pi k\dd k}{(2\pi)^2}}
=1
\nonumber\\
Z_4^\text{QFL}  &= 
\frac{\int_0^1 \frac{2\pi k\dd k}{(2\pi)^2}\ k^4}
{(2\pi \int_0^1 \frac{2\pi k\dd k}{(2\pi)^2})^2  \int_0^1 \frac{2\pi k\dd k}{(2\pi)^2}}
=\frac 43
\end{align}

\begin{figure}[t]
\hskip -1.75in
\includegraphics[height=1.6in]{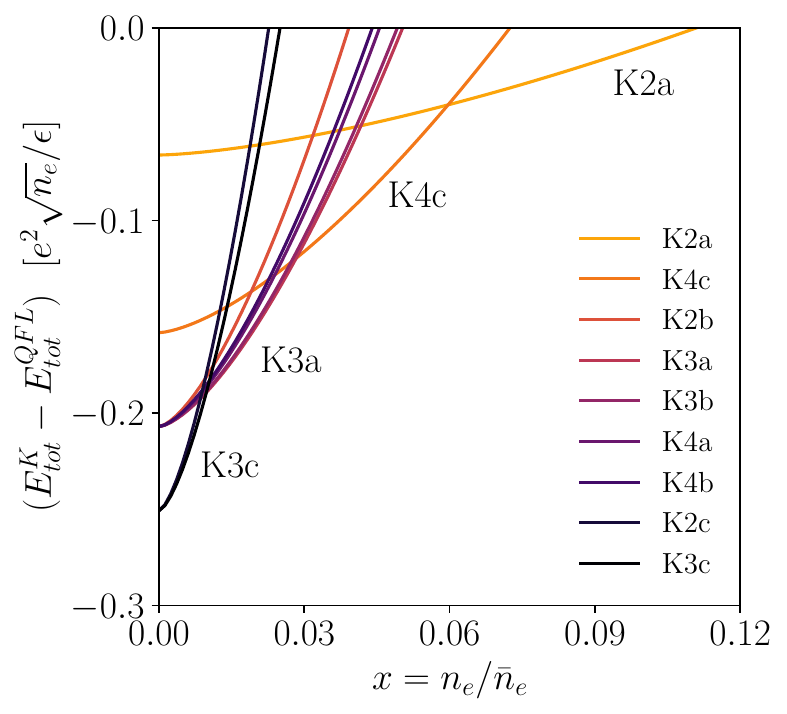}%
\includegraphics[height=1.6in]{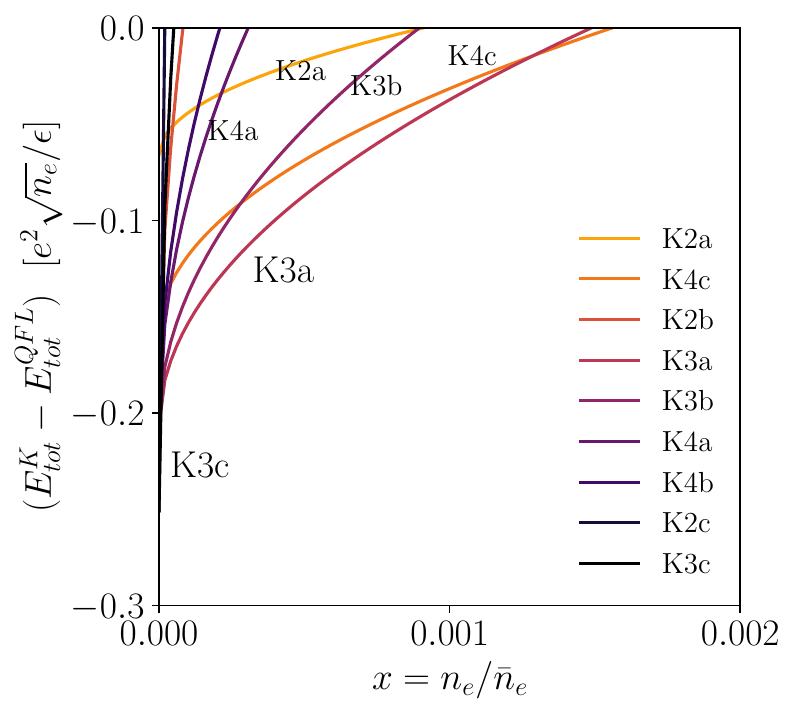}
\caption{
The normalized ground state energy
$\frac{E_\text{tot}^K(n_e)}{e^2\sqrt{n_e}/\eps}$, 
minus
the normalized ground state energy of quarter Fermi liquid
$\frac{E_\text{tot}^\text{QFL}(n_e)}{e^2\sqrt{n_e}/\eps}$, 
is plotted as a
function of normalized density $x=n_e/\bar n_e$.  \\
\textbf{Left}: For dispersion $\varepsilon\sim k^4$ (\ie $\ga=4$) and $x\in
[0,0.12]$. 
The tetralayer graphene at $\ga=4$ has $\bar n_e = 4.6 \times
10^{12}$cm$^{-2}$.  
\\ \textbf{Right}: For dispersion
$\varepsilon\sim k^2$ (\ie $\ga=2$) and $x\in [0,0.002]$,
The tetralayer
graphene at $\ga=2$ has $\bar n_e = 1.2 \times 10^{12}$cm$^{-2}$.\\
We see that for $\ga=4$ case, the chiral superconductivity appears around density
$10^{12}$cm$^{-2}$ (\ie have energies less than that of quarter Fermi liquids), while for $\ga=2$ case, the chiral superconductivity appears around
$0.02\times 10^{12}$cm$^{-2}$. 
The Wigner crystal was observed experimentally below $n_e \sim
0.5\times 10^{12}$cm$^{-2}$.
} \label{phase42} 
\end{figure}

To plot the energy of the various states as a function of electron density
$n_e$, it is helpful to measure $ E_\text{tot}^K(n_e)$  in the unit of $
\frac{e^2\sqrt{n_e}}{\eps}$.
We find the normalized energy to be
\begin{align}
\label{EtotNorm}
 \frac{E_\text{tot}^K(n_e)}{e^2\sqrt{n_e}/\eps }
&= 
 V + \frac{\eps c_\ga}{e^2} (2\pi)^{\gamma/2} (n_e)^{(\ga-1)/2} Z_\ga
\nonumber\\
&=
 V + x^{(\ga-1)/2} Z_\ga
\end{align}
where $x$ describes the total electron density in unit of $\bar n_e$:
\begin{align}
 n_e = x \bar n_e, \ \ \
\bar n_e = \frac{1}{(2\pi)^{\gamma/(\gamma-1)}} \Big( \frac{e^2}{\eps c_\ga} \Big)^{2/(\ga-1)}
\end{align}
To estimate $\bar n_e$ for the tetralayer graphene at $\ga=4$, we use the
band dispersion  $10$meV $= c_4 (0.1/a_0)^4$ where $a_0 = 0.246$nm $= 4.65 a_B$, 
%
giving $\bar n_e = 4.6 \times 10^{12}$ cm$^{-2}$.  To estimate $\bar n_e$ for
the tetralayer graphene at $\ga=2$, 
 we use
 band dispersion $30$meV $= c_2 (0.1/a_0)^2$,
%
giving $\bar n_e = 1.2 \times 10^{12}$ cm$^{-2}$.

In Fig. \ref{phase42}, we plot the normalized energy
$\frac{E_\text{tot}^K(n_e)}{e^2\sqrt{n_e}/\eps}$ (minus that for quarter Fermi
liquid) as a  function of normalized density $x=n_e/\bar n_e$, for $\ga=4$ and
for $\ga=2$ cases. We find  that for $\ga=4$ case, the chiral superconductivity
appears below a density $ n_e \sim 10^{12}$cm$^{-2}$ (\ie have energies less than that of
quarter Fermi liquid). Experimentally, chiral superconductivity was observed
around $n_e \sim 0.5  \times 10^{12}$cm$^{-2}$ and a Wigner crystal was observed
below $n_e \sim 0.5\times 10^{12}$cm$^{-2}$ \cite{HJ240815233,LJ230917436}. Our
$\ga=4$ result fits experimental result very well. For $\ga=2$ case, the chiral
superconductors has energies less than that of quarter Fermi liquid below a
density $n_e \sim 0.02\times 10^{12}$cm$^{-2}$, which do not lead to chiral
superconductivity, since at such a low density, Wigner crystal has lower energy.
Thus, a flat dispersion is very helpful for chiral superconductivity.

We remark that, for $\ga=2$ and $\ga=4$ cases, the energy of
$K_{2a}$ chiral superconductor in Fig. \ref{phase42} is obtained with a fixed wave function
described by
\begin{align*}
\bar K^+ = \begin{pmatrix}
    1 & 0 \\
    0 & 1 \\
\end{pmatrix}
, \ \ \ \
\bar K^- = \begin{pmatrix}
    0 & 1 \\
    1 & 0 \\
\end{pmatrix}
\end{align*}
At low densities, the energy
of $K_{2a}$ chiral superconductor can be lowered further by increasing
$\bar K^\pm$:
\begin{align*}
\bar K^+ = \begin{pmatrix}
    1+\del K & \del K \\
    \del K & 1+\del K \\
\end{pmatrix}
, \ 
\bar K^- = \begin{pmatrix}
    \del K & 1+\del K \\
    1+\del K & \del K \\
\end{pmatrix}
\end{align*}
This further minimized energy of $K_{2a}$ chiral superconductor is given by the
dashed-curve in Fig. \ref{phase42}. So at low densities, according to our
calculation, many different chiral superconductors have similar energies
including the $K_{2a}$ chiral superconductor. Further study is needed to
determine which one has the lowest energy.

All the superconducting states studied in the paper, $K_{2a}$, $K_{2b}$, \etc,
break the time-reversal and reflection symmetry in the orbital motion of
electrons, and thus carry magnetic moment.  The $K_{2a}$, $K_{2b}$, and
$K_{2c}$ superconducting states has two spin-valley components with a higher
density, and the other two spin-valley components with a lower density. The
$K_{4a}$, $K_{4b}$ and $K_{4c}$ superconducting states have all four
spin-valley components at equal density. The other superconducting state has
one spin-valley component with a highest density, and the other three
spin-valley components with lower densities.  

Because the densities in different spin-valley components have quantized ratios
and is very different between those chiral superconducting states and quarter
Fermi liquid, the zero-temperature transition between those $K$-matrix chiral
superconducting states and quarter Fermi liquid is first order.

\section{Chiral superconductors
with non-trivial topological orders}

\def\arraystretch{1.3}
\begin{table*}[tb] 
\caption{Topological properties of the chiral superconductors. The table lists 
the number $N_\text{top}$ of topological excitations (with no  $A_\mu$-flux \ie no vortitopologicalcity),
the charge-$e_{SF}$ condensation (which determine 
the minimal electromagnetic $A_\mu$-flux quantum),
chiral central charge $c$ (\ie the number of chiral edge modes),
the fractions for each spin-valley components $f_I$,
the average orbital angular momentum $\<L\>$ per electron,
and the type of topological order in the topological chiral  superconductor.}
\label{SFtablechiral} 
\centering
\begin{tabular}{|c|c|c|c|c|c|c|c|}
\hline
$K$-matrix & $N_\text{top}$  & $e_{SF}$ & $c$  & $f_I$ & $\<L\>$ & type of topological order\\
\hline
$K_{2a}$ \eqref{K2a} & 1 & $2$ & 1 & $\frac12,\frac12$ & $-\frac12$ & $ \begin{matrix} \text{$K^\text{top}=(1)$}, \
\vc e^\text{top} = (1)
\\[-1mm] \text{``spin''-triplet $p+\ii p$ superconductor } \\ 
\end{matrix} $  \\
\hline
$K_{2b}$ \eqref{K2b} & 3 & $2$ & 1 & $\frac12,\frac12$ & $-\frac32$ & $K^\text{top}=(3)$, 
$\vc e^\text{top} = (1)$
\\
\hline
$K_{2c}$ \eqref{K2c} & 5 & $2$ & 1 & $\frac12,\frac12$ & $-\frac52$ & $K^\text{top}=(5)$,
$\vc e^\text{top} = (1)$
\\
\hline
$K_{3a}$ \eqref{K3a} & 1 & $4$ & 0 & $\frac14,\frac12,\frac14$ & $1$ & $K^\text{top}=\begin{pmatrix}
-1 & 0 \\ 
0 & 1 \\
\end{pmatrix}$,
$\vc e^\text{top} = (1,1)$
\\
\hline
$K_{3b}$ \eqref{K3b} & 1 & $6$ & 0 & $\frac16,\frac12,\frac13$ &  $-1$ & $K^\text{top}=
\begin{pmatrix}
    -1 & 0 \\ 
    0 & 1 \\
\end{pmatrix}
$,
$\vc e^\text{top} = (-3,1)$
\\
\hline
$K_{3c}$ \eqref{K3c} & 10 & $4$ & 0 & $\frac14,\frac12,\frac14$ & $-1$ & $K^\text{top}=
\begin{pmatrix}
    -5 & 0 \\ 
    0 & 2 \\
\end{pmatrix}
$,
$\vc e^\text{top} = (9,4)$
\\
\hline
$K_{4a}$ \eqref{K4a} & 1 & $4$ & $-1$ & $\frac14,\frac14,\frac14,\frac14$ & $-\frac12$ & $K^\text{top}=
\begin{pmatrix}
-1 & 0 & 0 \\ 
0 & -1 & 0 \\ 
0 & 0 & 1 \\
\end{pmatrix}
$,
$\vc e^\text{top} = (1,7,5)$
\\
\hline
$K_{4b}$ \eqref{K4b} & 12 & $4$ & 3 & $\frac14,\frac14,\frac14,\frac14$ & $-\frac12$ & $K^\text{top}=
\begin{pmatrix}
-2 & 0 & 0 \\ 
0 & 3 & 0 \\ 
0 & 0 & -2 \\ 
\end{pmatrix}
$,
$\vc e^\text{top} = (2,3,2)$
\\
\hline
$K_{4c}$ \eqref{K4c} & 1 & $4$ & -1 & $\frac14,\frac14,\frac14,\frac14$ & $\frac12$ & $K^\text{top}=
\begin{pmatrix}
-1 & 0 & 0 \\ 
0 & -1 & 0 \\ 
0 & 0 & 1 \\ 
\end{pmatrix}
$,
$\vc e^\text{top} = (3,1,3)$
\\
\hline
\end{tabular}
\end{table*}

The above chiral superconductors described by  \eqref{PsibarK} usually have
non-trivial topological orders and are beyond BCS.  We now compute the
topological order and other properties of such chiral superconductors.
Following \Rf{ZHK8982,WZ9174,TW1328}, we start with the
effective Lagrangian for the chiral superconductor described by \eqref{PsibarK}:
\begin{align}
\label{Lsc}
\cL =&
 \frac{K_{IJ}}{4\pi} a_{I\mu}\partial_\nu  a_{J\lambda}\epsilon^{\mu\nu\lambda}
-
\frac{e_I}{2\pi}
A_\mu \partial_\nu a_{I \lambda } \epsilon^{\mu\nu\lambda}
\nonumber\\
& - a_{I0} l_I \del(\vc x),\nonumber\\
 \vc e^\top &= (1,1,\cdots),
\end{align}
where
\begin{align}
 K = \bar K^+ - \bar K^- 
\end{align}
and $J_\mu^I = \frac{1}{2\pi}
\partial_\nu a_{I \lambda } \epsilon^{\mu\nu\lambda}$
are the density and current of $I^\text{th}$ spin-valley component.
The  above effective theory is a compact $U(1)$ Chern-Simons theory with
integral quantized $U(1)$ charges $l_I$.  The $K$ has a zero eigenvalue
which gives rise to the gapless superfluid mode.

It is convenient to choose an integral basis to make  $K$ to have the following block form
\begin{align}
\label{blockdiag}
\tl K =
\begin{pmatrix}
 K^\text{top} & \vc 0^\top \\
 \vc 0 & 0 \\
\end{pmatrix}.
\end{align}
Such an integral basis always exists.  First $K$ can be written as
$K = U D W$, where $U$ and $W$ are unimodular integral matrices, and
$D$ is the Smith normal form of $K$, which is a diagonal matrix.
Since $K$ has an zero eigenvalue, diagonal of $D$ has a form
$(D_{11},D_{22},\cdots,D_{\kappa\kappa}=0)$, where $\kappa$ is the dimension of $K$.
Now we introduce $\tl a_{I \mu}$ via
\begin{align}
 a_{I \mu} = (W^{-1})_{IJ} \tl a_{J \mu},
\end{align}
and write the  effective Lagrangian \eqref{Lsc} in terms of $\tl a_{I \mu}$:
\begin{align}
\label{LCS}
\cL &=
 \frac{\tl K_{IJ}}{4\pi} \tl a_{I\mu}\partial_\nu  
 \tl a_{J\lambda}\epsilon^{\mu\nu\lambda}
-
\frac{\tl e_I}{2\pi}
A_\mu \partial_\nu \tl a_{I \lambda } \epsilon^{\mu\nu\lambda}
 - \tl a_{I0} \tl l_I \del(\vc x),
\nonumber\\
\tl K 
&= 
(W^\top)^{-1} K W^{-1},\ 
\tl {\vc e} = (W^\top)^{-1} \vc e = (\vc e^\text{top}, e_{SF})
\end{align}
We see that
\begin{align}
 \tl K = (W^\top)^{-1} U D.
\end{align}
Such a matrix is symmetric and integral.  Since $D_{I\kappa}=0$, the last column of
$\tl K$ is zero. The last row of $\tl K$ is also
zero since $\tl K$ is symmetric.

When there is a gapless mode, we must include Maxwell terms as the new leading
order contribution in the superfluid sector.  Most generally, this includes terms $g_{IJ} \tl
f_{I,\mu\nu} \tl f_{J}^{\mu\nu}$, where $\tl f_{I,\mu\nu} = \partial_\mu \tl
a_{I,\nu} - \partial_\nu \tl a_{I,\mu}$.  We further decompose $g$ into $g^\text{top}$ purely in the 
topological sector, $\vc g$ coupling the topological and superfluid sectors, and 
$g_{SF}$ purely in the superfluid sector.  

Fourier transforming and writing $p^{\mu\nu} =
\eps^{\mu\nu\lambda}p_\lambda$, the  effective Lagrangian \eqref{LCS}
becomes at leading order
\begin{align}
\label{LcSC}
    &\cL =  
     \frac{\tl{\vc e}}{2\pi} A_\mu p^{\mu\nu} \tl {\vc a}_\nu
    - \tl {\vc l} \cdot \tl {\vc a} \; \delta(\vc x)
    \\ 
    & 
    + \tl {\vc a}_\mu \begin{pmatrix}
        \tfrac{K^\text{top}}{4\pi} \; p^{\mu\nu} + g^\text{top} \; {p^{\mu}}_\sigma p^{\sigma\nu} & \vc g \; {p^{\mu}}_\sigma p^{\sigma\nu} \\
        \vc g \; {p^{\mu}}_\sigma p^{\sigma\nu} & 
        g_{SF} \; {p^{\mu}}_\sigma p^{\sigma\nu} & 
    \end{pmatrix} \tl {\vc a}_\nu
    \nonumber
\end{align}

The splitting of the basis into a superfluid and topological sector is not 
well-defined, and in particular we are free to add the superfluid mode to any 
topological basis vector (say with coefficients $t_{I'}$, where primed indices
go up to $\kappa - 1$) and preserve the form 
of $K^\text{top}$.  Under such a transformation, the gauge field labels 
transform as $a_{SF} \rightarrow a_{SF} - t_{I'} a_{I'}$, 
the charge vector  transforms as $e_{I'} \rightarrow e_{I'} + t_{I'} e_{SF}$, 
the excitation label  transforms as $\tl l_{I'} \rightarrow \tl l_{I'} + t_{I'} \tl l_{SF}$, 
and the $g_{I'}$ transform 
as $g_{I'} + t_{I'} g_{SF}$.  Therefore we can block-diagonalize the $g_{IJ}$
by choosing $t_{I'} = -g_{I'}/g_{SF}$.  This basis is no longer charge quantized, 
but the superfluid mode is now decoupled from the topological sector.
\begin{align}
\label{LcSC2}
    \cL = & \; \tl {\vc a}_\mu \begin{pmatrix}
        \tfrac{1}{4\pi} K^\text{top} \; p^{\mu\nu} & \vc 0 \\
        \vc 0^\top  & 
        g_{SF} \; {p^{\mu}}_\sigma p^{\sigma\nu} & 
    \end{pmatrix} \tl {\vc a}_\nu
    + \frac{\tl{\vc e}}{2\pi} A_\mu \hat p^{\mu\nu} \tl {\vc a}_\nu
    \nonumber \\
    &- \tl {\vc l} \cdot \tl {\vc a} \; \delta(\vc x)
\end{align}
We have also dropped
$g^\text{top}$ as a subleading contribution that will not contribute at any
order relevant in the transport calculations which follow.  We stress that now $\tl {\vc e} = (\tl{\vc
e}^\text{top}, e_{SF})$ with $\tl{\vc e}^\text{top}$ no longer quantized as integers, but $e_{SF}$ is unchanged and is still an integer.

We see that the electron current is given by 
\begin{align}
\label{Jmu}
 J^\mu &= \frac{\tl {\vc e}}{2\pi}\cdot \partial_\nu \tl {\vc a}_{\lambda }
\epsilon^{\mu\nu\lambda}
.
\end{align}

The gauge field $ \tl a_{SF \mu}$ in the last component of $\tl {\vc a}$ is gapless and describes the gapless
superfluid mode of the chiral superconductor.  The other gauge fields $ \tl
a_{I \mu}$ are gapped and describe the topological sector of the chiral
superconductor with topological order.

The chiral superconductor behaves like a usual superconductor
stacked with a topological sector that may have a non-zero Hall conductance.
To understand the responses, we integrate out the dynamical gauge fields. The new Lagrangian describing a purely electromagnetic response is formally
\begin{align}
    \cL^{EM}&\; =  
\frac{1}{4}p^{\mu\lambda}A_\lambda \frac{\tl {e}^\text{top}_I}{2\pi} \left(\tfrac{1}{4\pi} K^\text{top}p_{\mu\nu} + i\epsilon\right)^{-1}_{IJ}
    \frac{\tl{e}^\text{top}_J}{2\pi} p^{\nu\rho} A_\rho \nonumber \\
    & + \frac{ e^2_{SF}}{4(2\pi)^2}p^{\mu\lambda}A_\lambda\left( g_{SF} \; {\hat\partial^{\mu}}_\sigma \hat\partial^{\sigma\nu} + i\eps\right)^{-1}p^{\nu\rho} A_\rho
\end{align}
where we added an $i\epsilon$ identity term to the quadratic-in-$\tl {\vc a}$ term to make the integrals converge. 

In the first term, a $p^{\mu\lambda}$ and $(p_{\mu\nu})^{-1}$ cancel each other without poles, so we can simplify to find
the transverse conductance
\begin{align}
    \cL^{EM}_\text{top} = \sigma \epsilon^{\lambda \mu \nu} A_\lambda \partial_\mu A_\nu \nonumber \\
    \sigma = \frac{1}{4\pi} \tl e^\text{top}_I \left(K^\text{top}\right)^{-1}_{IJ} \tl e^\text{top}_J
\end{align}
To emphasize, this Hall conductance is not quantized because the $\tl e^\text{top}$ chosen to block diagonalize $g_{IJ}$ are not integer.

Some care must be taken contracting the momenta in the superfluid term.  We work in Lorentz signature $(-,+,+)$, Fourier transform, and fix the gauge to $a_0 = 0$.  In a basis $\{\tl a_t, \tl a_\ell\} = \{ (0, -p_y, p_x), (0, p_x, p_y) \}$ labeling the transverse and longitudinal components respectively, we obtain

\begin{align}
    \hat p^{\mu\lambda}\left( {\hat p^\mu}_\sigma \hat p^{\sigma\nu} + i\epsilon \right)^{-1} p^{\nu\rho} = 
    \begin{pmatrix}
        \frac{\omega^2 + p_x^2 + p_y^2}{\omega^2 + p_x^2 + p_y^2 + i\eps} & 0 \\
        0 & \frac{\omega^2 + p_x^2 + p_y^2}{\omega^2 - p_x^2 - p_y^2 + i\eps} 
    \end{pmatrix} \nonumber
\end{align}
The term in the transverse component has no pole, so we can take $\eps$ to zero and replace it by $1$.
The lowest order term in the EM Lagrangian is obtained by contracting this with $A_t$ and $A_\ell$ in the same basis.  Some intuition for this basis can be gained by Fourier transforming back to position space.  We have $A_t(\vc x) = (0, -\partial_y\psi_t, \partial_x\psi_t) $ and $A_\ell (\vc x) = (0, \partial_x\psi_\ell, \partial_y\psi_\ell) $ where the $\psi$ are scalar functions of position and time.  The longitudinal part describes pure electric field due to vanishing curl.  The transverse part describes a nonzero background magnetic field if $\nabla^2 \psi_t \neq 0$.  Since we are interested in the case of zero magnetic field, we deduce that $\psi_t$ is a solution of the Laplace equation.

Returning to momentum space, a general vector potential $A_\mu$ can be separated into longitudinal and transverse parts via the projectors
\begin{align}
     A_\ell^i = \frac{p^i p_j}{p_x^2 + p_y^2}  A^j, \quad  A_t^i = \frac{\epsilon^{i \ell}p_\ell \epsilon_{jk} p^k}{p_x^2 + p_y^2}  A^j
\end{align}
where we use Latin indices to emphasize that $A$ only has $x$ and $y$ vector potential components and $A_0=0$.

We can now write the lowest order part of the EM Lagrangian

\begin{align}
    \cL^{EM}_{SF} = \frac{e_{SF}^2}{4(2\pi)^2 g_{SF}}& \left( 
     A^\mu_t  A_{t,\mu}
     + \frac{\omega^2 + p_x^2 + p_y^2}{\omega^2 - p_x^2 - p_y^2 + i\eps}  A_\ell^\mu  A_{\ell,\mu} \right)
\end{align}
The supercurrent is
\begin{align}
    J_{SF}^\mu &\;= \frac{e_{SF}^2}{2(2\pi)^2 g_{SF}}\left( \frac{\epsilon^{\mu i}p_i \epsilon_{jk} p^k}{p_x^2 + p_y^2}\right. \nonumber \\
    &\ \ \ \ \ 
    + \left.\frac{\omega^2 + p_x^2 + p_y^2}{\omega^2 - p_x^2 - p_y^2 + i\eps} \ \frac{p^\mu p_j}{p_x^2 + p_y^2} \right)  A^j
\end{align}
This is the London equation for superconductors, where the gauge of $A$ has been fixed by $A_0 = 0$.  

Thus the total response current is
\begin{align}
    J^i & \;= \si \eps^{i j} \omega  A_j 
    +  \frac{e_{SF}^2}{2(2\pi)^2 g_{SF}}\left( \frac{\epsilon^{i \ell}p_\ell \epsilon_{jk} p^k}{p_x^2 + p_y^2} \right. \nonumber \\
    &\ \ \ \ \ 
    + \left.\frac{\omega^2 + p_x^2 + p_y^2}{\omega^2 - p_x^2 - p_y^2 + i\eps} \ \frac{p^i p_j}{p_x^2 + p_y^2}  \right)  A^j
\end{align}
where $\omega  A_j$ is the electric field, and
as before we use Latin indices to indicate spatial degrees of freedom.

To obtain the topological order in the chiral superconductor, let us still use
the basis in the effective Lagrangian \eqref{LcSC2}, where the superfluid sector and the topological sector
completely decouple. We need to study its topological excitations, which are
labeled by vectors $(\tl l_{I'},\tl l_{SF})$, $I'=1,\cdots,\ka - 1$. In this basis,
$\tl l_{SF}$ is still an integer, but $\tl l_{I'}$ may not be integer. However,
when $\tl l_{SF} = 0$, $\tl l_{I'}$ are quantized as integers.

The term $ -\tl a_{I0} \tl l_I \del(\vc x) -\tl a_{SF, 0} \tl
l_{SF} \del(\vc x) $ in the effective Lagrangian describes such a topological
excitation located at $\vc x=0$.  
From the equation of motion for $\tl a_{SF,0}$,
\begin{align}
-
\frac{e_{SF}}{2\pi}
\partial_i A_j \epsilon^{ij}
= \tl l_{SF}\delta(\vc x),
\end{align}
we obtain
\begin{align}
\frac{1}{2\pi} \partial_i   A_j \epsilon^{ij} = - e_{SF}^{-1}\tl l_{SF}\del(\vc x)
.
\end{align}
We see that if $ \tl l_{SF} \neq 0$, the excitation corresponds to a vortex in
the chiral superconductor \cite{WZ9174,TW1328}.  Such a vortex carries a $
e_{SF}^{-1}$ unit of $A_\mu$-flux:
\begin{align}
 A_\mu\text{-flux} = 
-2\pi
e_{SF}^{-1}
.
\end{align}
Thus, the chiral superconductor has a charge-$e_{SF}$ condensation.

When $\tl l_{SF}=0$, the excitations labeled by integers $\tl {\vc l}^{\text{top},\top}= (\tl l_{I'})$
are finite energy excitations in the chiral superconductor (\ie not vortices).
This kind of excitations 
carry fractional electric charges
\begin{align}
q_{\tl{\vc l}^\text{top}} = 
  \tl{\vc e}_1^{\text{top}\top} (K^\text{top})^{-1}\tl{\vc l}^\text{top}. 
\end{align}
which are not quantized.
They also have statistics and mutual statistics
\begin{align}
 s_{\tl{\vc l}^\text{top}} &= \frac12 \tl{\vc l}^{\text{top}\top} (K^\text{top})^{-1}\tl{\vc l}^\text{top},
 \nonumber\\
 s_{\tl{\vc l}_1^\text{top},\tl{\vc l}_2^\text{top}} &=  \tl{\vc l}_1^{\text{top}\top} (K^\text{top})^{-1}\tl{\vc l}_2^\text{top}. 
\end{align}
which are fully determined by the $(K^\text{top})$-matrix.  The $S$ and $T$ matrices are
defined as
\begin{align}
\label{STmatK}
 \mathcal{S}_{ij} = \tfrac{1}{\sqrt{N_\mathrm{top}}}e^{-2\pi \ii s_{ij}}, \quad
 \mathcal{T}_{ij} = \delta_{ij} e^{\ii 2\pi s_{i}}
\end{align}
which describes the topological order in the chiral superconductors.

Beside the $\mathcal{S},\mathcal{T}$-matrices, there is another topological
invariant to describe topological order, which is the chiral central charge $c$,
defined as the number of positive eigenvalues minus the number of negative
eigenvalues of $K^\text{top}$.  Physically, the chiral central charge is the  number
of right-moving edge modes minus the  number of left-moving edge modes, which
can be measure via thermal Hall conductance \cite{KF9732}.

For the $K_{2a}$ chiral superconductor \eqref{K2a}, its topological order is
described by
\begin{align}
\label{tK2a}
 K^\text{top}_{2a} &= \begin{pmatrix}
1 \\ 
\end{pmatrix}
, \ \ \text{det}(K^\text{top}_{2a}) = 1, \ \ c=1,
\nonumber\\
e_{SF} &= 2,\ \ \ \ {\vc e^\text{top}} = (1)
\end{align}
Since $\text{det}(K^\text{top}_{2a}) = 1$ and $c =1$, the $K_{2a}$ chiral
superconductor is a topological superconductor, that has no bulk anyons (since
$\text{det}(K^\text{top}_{2a}) = 1$).  The topological nature of the superconductor is
characterized by its single chiral edge mode (since $c=1$).  Such kind of
topological order without non-trivial bulk anyons is called invertible
topological order.  Since the diagonal of $\tl K_{2a}$ contains odd integers,
the invertible topological order is a fermionic invertible topological order.
Such a fermionic invertible topological order is characterized by the chiral
centetral charge $c=1$ plus the following $\mathcal{S},\mathcal{T}$-matrices
\begin{align}
\label{fermion_ST}
\mathcal{S}_{\text{ferm}} = \frac{1}{\sqrt{2}} \begin{pmatrix}
	1 & 1 \\
	1 & 1
\end{pmatrix}, \quad
\mathcal{T}_{\text{ferm}} =  \begin{pmatrix}
	1 & 0 \\
	0 & -1 
\end{pmatrix}
\end{align}
We point out that chiral central charge $c=0$ plus the above
$\mathcal{S},\mathcal{T}$-matrices will describe a trivial fermion product
state. 

$|e_{SF}|= 2$ indicates that the superconductor has a charge-2 condensation.
Such a chiral superconductor is a BCS ``spin''-triplet $p+ip$-wave
superconductor.

For the $K_{2b}$ chiral superconductor \eqref{K2b}, its topological order is
described by
\begin{align}
\label{tK2b}
 K^\text{top}_{2b} &= \begin{pmatrix}
3 \\ 
\end{pmatrix}
, \ \ \text{det}(K^\text{top}_{2a}) = 3, \ \ c=1,
\nonumber\\
e_{SF} &= 2,\ \ \ \ {\vc e^\text{top}} = (1)
\end{align}
Since $\text{det}(K^\text{top}_{2b}) = 3$ and $c =1$, the $K_{2b}$ chiral
superconductor is a topological superconductor, that has non-trivial bulk
anyons.  Such a chiral superconductor is a beyond-BCS superconductor.  The
topological properties of other chiral superconductors are summaries in Table
\ref{SFtablechiral}.

\section{Pfaffian chiral superconductors}

\begin{figure}[t]
\hskip -1.75in
\includegraphics[height=1.6in]{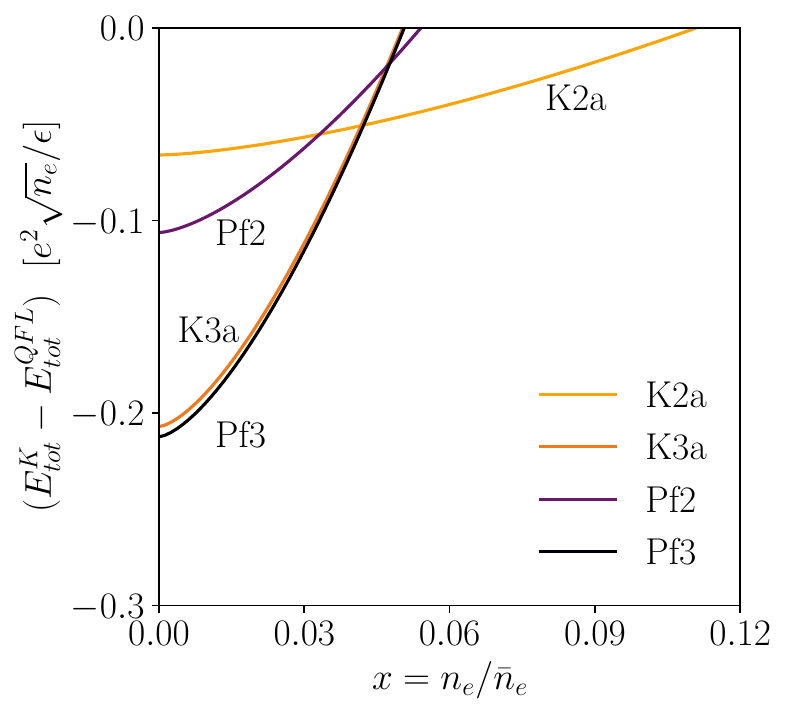}%
\includegraphics[height=1.6in]{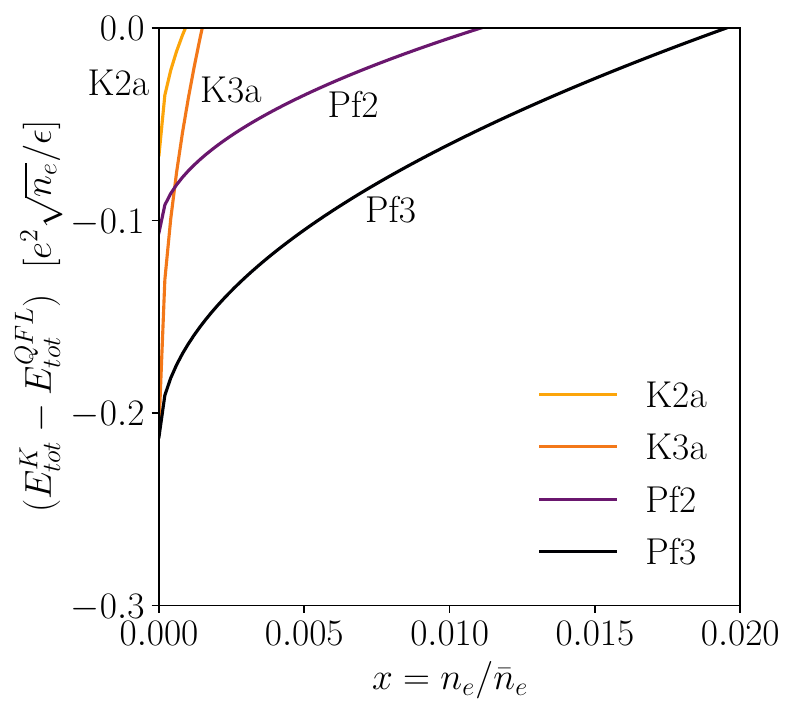}
\caption{
The normalized ground state energy
$\frac{E_\text{tot}^K(n_e)}{e^2\sqrt{n_e}/\eps}$ for the Pfaffian
superconductors \eqref{Pfm}, Pf2 and Pf3, with $m=2,3$, minus the normalized
ground state energy of quarter Fermi liquid
$\frac{E_\text{tot}^\text{QFL}(n_e)}{e^2\sqrt{n_e}/\eps}$, is plotted as a
function of normalized density $x=n_e/\bar n_e$.  \\ \textbf{Left}: For
dispersion $\varepsilon\sim k^4$ (\ie $\ga=4$) and $x\in [0,0.12]$.  The
tetralayer graphene at $\ga=4$ has $\bar n_e = 8.5 \times 10^{12}$cm$^{-2}$.
\\ \textbf{Right}: For dispersion $\varepsilon\sim k^2$ (\ie $\ga=2$) and $x\in
[0,0.02]$, The tetralayer graphene at $\ga=2$ has $\bar n_e = 9.8 \times
10^{12}$cm$^{-2}$.\\ We see that for $\ga=4$ case, the Pfaffian superconductors
appear around density $0.5\times 10^{12}$cm$^{-2}$, while for $\ga=2$ case, the Pfaffian
superconductors appear around $0.2\times 10^{12}$cm$^{-2}$.  The Wigner crystal
was observed experimentally below $n_e \sim 0.5\times 10^{12}$cm$^{-2}$.  }
\label{phase42Pf} 
\end{figure}

The many-body wave function of the chiral superconductors discussed above has a
product form -- a product of factors of form $f(z_i-z_j)$ and $g(z_i)$.  We can
compute the potential energy of these wave functions easily, allowing us to
propose a new class of superconducting states beyond the BCS mechanism.
 
In this section, we discuss another class of chiral superconductors -- called
secondary chiral superconductors -- whose wave function is more complicated
than the simple product form (\ie contains sums).
In particular, we consider the following Pfaffian wave function
\begin{align}
\label{Pfm}
 \Psi(\{z_i\}) &= \cA( \frac 1{z_1-z_2}
\frac 1{z_3-z_4} \cdots) 
\nonumber\\
&\ \ \ \
\prod_{i<j} |z_i-z_j|^m 
\ee^{-\frac{1}{4l_b} \sum_i |z_i|^2}
\end{align}
where $\cA$ generates an anti-symmetrization operation.
Because $\cA( \frac 1{z_1-z_2}
\frac 1{z_3-z_4} \cdots) $ is a  Pfaffian
\begin{align}
 \cA( \frac 1{z_1-z_2}
\frac 1{z_3-z_4} \cdots) = \text{Pf}(\frac1{z_i-z_j})
\end{align}
it is still possible to compute the average Coulomb energy of the Pfaffian wave
function.

The energy for the Pfaffian superconductor \eqref{Pfm} has a form
\eqref{EtotNorm}.  For Pf2 state in \eqref{Pfm} with $m=2$, we have
\begin{align}
 \label{Pf2}
V & = -1.6108, \ \ 
Z_2  \sim 1+2-0.99, \ \ 
Z_4  \sim \frac43+8+0.46. 
\end{align}
For Pf3 state in \eqref{Pfm} with $m=3$, we have
\begin{align}
 \label{Pf3}
V & = -1.7169, \ \
Z_2  \sim 1+3-1.48, \ \ 
Z_4  \sim \frac43+ 18+0.65. 
\end{align}
The kinetic energy of the Pfaffian superconductor \eqref{Pfm} is difficult to
calculate.  We calculate the kinetic energy of the following state instead
\begin{align}
\prod_{i<j} |z_i-z_j|^m 
\ee^{-\frac{1}{4l_b} \sum_i |z_i|^2}.
\end{align}
We then increase the resulting $Z_2$ and $Z_4$ by those of quarter Fermi
liquid, $Z_2^\text{QFL}=1$ and $Z_4^\text{QFL}=\frac43$, to simulate the effect
of anti-symmetrization.  We use the resulting $Z_2$ and $Z_4$ to estimate the
kinetic energy of Pfaffian superconductor \eqref{Pfm}.

From Fig. \ref{phase42Pf}, we see that for dispersion $\veps \sim k^4$, the
Pfaffian superconductor appears around $0.5\times 10^{12}$cm$^{-2}$, while for dispersion
$\veps \sim k^2$, the Pfaffian superconductors appear around $0.2\times
10^{12}$cm$^{-2}$.  The Wigner crystal was observed experimentally below $n_e
\sim 0.5\times 10^{12}$cm$^{-2}$.  This suggests that Pfaffian superconductors
are not likely to appear for dispersion $\veps \sim k^2$.  For  dispersion $\veps \sim k^4$,
the Pfaffian superconductors have similar energies as the other chiral
superconductors discussed before.

We note that the Pfaffian superconductors are fully spin-valley polarized.  So
fully spin-valley polarized chiral superconductor, as realized by the Pfaffian
superconductors, are also quite possible.  We expect the Pfaffian
superconductors to belong to the same phase as the spinless $p+\ii p$ BCS
superconductor \cite{RG0067}.  Thus the  Pfaffian superconductors are topological
superconductors.

\section{Anyon superconductors}

Recently, fractional quantum anomalous Hall (FQAH) states were discovered in
\Rf{CX230408470,PX230802657,XL230806177}.  It was usually stressed that FQAH
states can appear at a zero magnetic field.  Here, we stress that FQAH
states are special because they are realized in a periodic potential.  If we
add electrons/holes to such lattice FQAH states, we will obtain an anyon gas
hopping in a background triangle lattice.  Such an anyon gas in triangle
lattice has been studied in \Rf{TW1328}, where anyon superconductivity with
intrinsic topological order was discovered.  However, \Rf{TW1328} only studied
certain possible anyon superconducting states.  In this section, we will
explore other possible anyon superconducting states which may be simpler.  We
hope among those possible anyon superconducting states, some of them can be
realized in neighborhood of FQAH states, when the periodic potential for anyons
is strong enough.  As stressed in \Rf{TW1328}, the anyon superconductivity near
FQAH phases is actually induced by strong Coulomb repulsion, and is a special
case of chiral superconductors discussed in the previous sections.

The doped FQAH states are also independently studied in \Rf{SS240920567}
recently. There, the lattice effects are more explicitly explored, which allows some other states
besides the Abelian anyon superfluids to be studied.

\begin{figure}[t]
\hskip -0.4in
\includegraphics[align=c,height=1.5in]{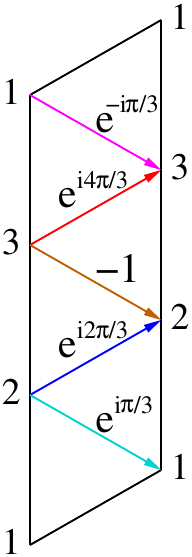}
\hskip -0.5in
$
\begin{matrix}
\includegraphics[align=c,height=0.6in]{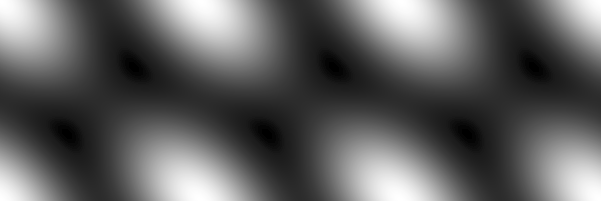}\\
\\
\includegraphics[align=c,height=0.6in]{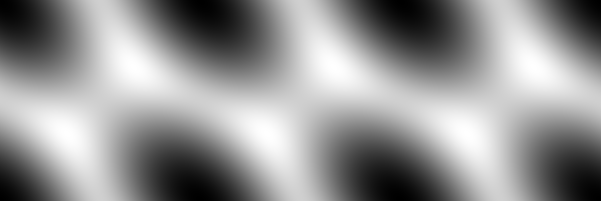}\\
\end{matrix}
$
\hskip 0.9in
\includegraphics[align=c,height=1.5in]{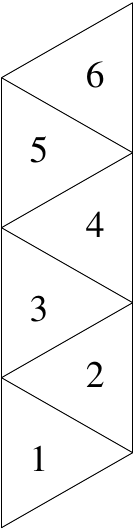}
\\
\caption{(left) The magnetic unit cell, where anyons have lowest energy at the
site of triangle lattice. The anyon hopping on a triangle lattice has hopping
amplitude $-t_{ij}$ whose phase is $0$ on black links, and on colored link as
indicated above.  The resulting lowest band in the magnetic Brillouin zone is
given by the middle-top graph, which has six minima (the black regions between
three white regions).  (right) The magnetic unit cell, where the anyons have
lowest energy at the centers of the triangles. The resulting lowest band has
three minimal points (the middle-bottom graph).  } \label{anyonhop} 
\end{figure}

Let us consider a filling fraction $\nu=\frac13$
FQAH state\footnote{The discussions in this section also apply to $\nu=2/3$  FQAH
state which can be viewed as $\nu=\frac13$ FQAH state of holes.} realized in a
2-dimension material with a moir\'e pattern.  The  moir\'e pattern usually form
a triangle lattice.  The FQAH state comes from a $\frac13$ filled flat Chern
band of Chern number $1$.  Thus the $\nu=\frac13$ FQAH state has an electron
density of $\frac13$ electron per moir\'e unit cell.

If we change the electron density, the added electrons will form an anyon gas
on the triangle lattice.  The fractional statistics of the anyon is
$\theta_a=\frac{\pi}{3}$.  We first assume that the anyons have lower energy at
the sites of the triangle lattice and the hopping of the anyons can be
described by a tight binding model on triangle lattice.

We note that an electron in the Chern band behaves like a $2\pi$ flux to the
anyon.  Thus, the tight binding model of anyon hopping contains $\frac{2\pi}{3}$
flux per moir\'e unit cell, making the hopping amplitudes complex as given by 
Fig. \ref{anyonhop}(left).  The resulting
anyon band is given by Fig. \ref{anyonhop}(right), which contain six minimal
points.  Therefore we have six species of anyons at low energies.

If we assume, instead, the anyons have lower energy at the centers of the
triangles, we find that the resulting anyon lowest band has three minimal points.
In this second case, we have three species of anyons at low energies.

\subsection{Anyon superconductivity for six species of anyons}

The six-species case was
discussed in \Rf{TW1328}, where $\theta_a=\frac{\pi}{3}$ anyon is viewed as a
fermion attached to $\frac{2\pi}{3}+2\pi = \frac{8\pi}{3}$ flux.  
The $2\pi \frac43 $ flux per fermion was then smeared into an uniform effective ``magnetic''
field.  This changes the anyon gas to a fermion gas of six species of fermions
with a total filling fraction $\tl \nu =\frac34$.  Such a fermion gas can lead to
several possible topological superconducting states.

In this paper, we will use a different mean-field approach, which leads to
different possible anyon superconducting states. We view the
$\theta_a=\frac{\pi}{3}$ anyon as a boson attached to $\frac{2\pi}{3}$ flux.  We
then smear the $2\pi \frac13 $ flux per boson into an uniform effective
``magnetic'' field.  This changes the anyon gas to a boson gas of six species
of bosons with a total filling fraction $\tl \nu =3$.  

To obtain the above results formally, we start with the effective
Lagrangian for the $\nu=\frac13$ FQH state of electrons
\begin{align}
\label{Lbosons}
\cL &=
\frac{3}{4\pi} \tl a_\mu\partial_\nu \tl a_\lambda\epsilon^{\mu\nu\lambda}
-\frac{1}{2\pi} A_\mu\partial_\nu \tl a_\lambda\epsilon^{\mu\nu\lambda}
+\ii \phi^\dag_I(\partial_0+\ii \tl a_0)\phi_I
\nonumber \\
&-\frac{1}{2m}|(\partial_i+\ii \tl a_i)\phi_I|^2+
\text{Coulomb interaction}
\end{align}
where $A_\mu$ is the external electromagnetic field, $J_\mu=\frac{1}{2\pi}
\partial_\nu \tl a_\lambda\epsilon^{\mu\nu\lambda}$ is the electron current, and
$\phi_I$, $I=1,\cdots,6$, are bosonic fields carrying unit of $\tl a_\mu$ charge.
Those bosonic fields describe the six species of anyon excitations.  

The six species of bosons form a FQH state described by $K$-matrix: 
\begin{align}
\Big[
 \prod_{I=1}^6 \prod_{i<j} (z_i^I-z_j^I)^{K_{II}}
 \prod_{I<J} \prod_{i,j} (z_i^I-z_j^J)^{K_{IJ}} 
\Big]
\ee^{\frac14 \sum |z_i^I|^2},
\end{align}
where $I,J=1,\cdots,6$ and $K$ is a $6\times 6$ integer matrix with even
diagonal elements.  
Such a $K$-matrix state, plus its parent electron $\nu=\frac13$ FQH
state, is described by the following total effective theory 
\begin{align}
\label{LKaSF}
&\cL=
\frac{3}{4\pi} \tl a_\mu\partial_\nu \tl a_\lambda\epsilon^{\mu\nu\lambda}
-\frac{1}{2\pi} A_\mu\partial_\nu \tl a_\lambda\epsilon^{\mu\nu\lambda}
-
\frac{q_I}{2\pi}
\tl a_\mu 
\partial_\nu a_{I \lambda } \epsilon^{\mu\nu\lambda}
\nonumber\\
&
+
\frac{K_{IJ}}{4\pi}  a_{I\mu}\partial_\nu 
 a_{J\lambda}\epsilon^{\mu\nu\lambda},
\ \ \ \ q_I = 1, \ \ I,J=1,\cdots,6.
\end{align}
where $\frac{1}{2\pi} \partial_\nu \tl a_{I\lambda} \epsilon^{\mu\nu\lambda}$ is
the current of $I^\text{th}$ anyons.  
This  effective theory 
has been discussed before (see \eqref{Lsc} and setting $k_f=3$),
which is the effective theory for anyon superconductor.

There are many $K$-matrices that can give rise to
 anyon superconductor.  To determine which
$K$-matrices are more likely, we roughly estimate the energy of each
$K$-matrix quantum Hall state.  We note that $K_{IJ}$ is the order of zeros in
the wave function as a species-$I$ anyon approach a species-$J$ anyon. For the potential energy, we do not have the potential energy fitting for the anyon-anyon interaction. However, we can make a crude ansatz based on the $b/(K_{IJ} + a)$ form we obtained for the electron-electron interactions, given as
\begin{align}
V_{IJ} =
\frac{1}{K_{IJ}+1}
\end{align}
such that $V_{IJ}$ represents the interaction energy between a species-$I$ anyon
and a species-$J$ anyon. The core principle of higher order of zeros giving lower Coulomb energy is therefore maintained. Since $\nu_I = (K^{-1})_{IJ} q_J$ is proportional to
the density of the species-$I$ anyons, we can estimate
the total energy of a $K$-matrix quantum Hall state as
\begin{align}
\label{Etot1}
 E_\text{tot}(K) = \nu_I V_{IJ} \nu_J
\end{align}
We remark that the kinetic energy of anyons is ignored in the above estimate.

We find a few $K$-matrices whose energies are low and close, and all the
species of anyons have non-zero positive densities.  The first one is
\begin{align}
 \label{K6A}
K_{6A} &=
\begin{pmatrix}
 0&  0&  0&  0&  1&  1\\
 0&  0&  0&  1&  0&  1\\
 0&  0&  0&  1&  1&  0\\
 0&  1&  1&  0&  0&  0\\
 1&  0&  1&  0&  0&  0\\
 1&  1&  0&  0&  0&  0\\
\end{pmatrix} ;
\ \ \
\text{det}(K_{6A}) = -4,
\end{align}
which in fact has the lowest energy. A similar $K$-matrix which also has the lowest energy is
\begin{align}
 \label{K6Ap}
K_{6A'} &=
\begin{pmatrix}
 0& 1& 1& 0& 0& 0\\
 1& 0& 1& 0& 0& 0\\
 1& 1& 0& 0& 0& 0\\
 0& 0& 0& 0& 1& 1\\
 0& 0& 0& 1& 0& 1\\
 0& 0& 0& 1& 1& 0\\
\end{pmatrix} ;
\ \ \
\text{det}(K_{6A'}) = 4.
\end{align} 
Other examples of $K$-matrices satisfying the superfluidity conditions with low energy are
\begin{align}
 \label{K6B}
K_{6B} &=
\begin{pmatrix}
 0& 0& 0& 1& 2& 1\\
 0& 0& 1& 0& 0& 0\\
 0& 1& 0& 0& 0& 0\\
 1& 0& 0& 0& 2& 1\\
 2& 0& 0& 2& 0& 0\\
 1& 0& 0& 1& 0& 2\\
\end{pmatrix} ;
\ \ \
\text{det}(K_{6B}) = -16 , 
\end{align}
\begin{align}
 \label{K6C}
K_{6C} &=
\begin{pmatrix}
 0& 0& 0& 2& 2& 0\\
 0& 0& 1& 0& 0& 0\\
 0& 1& 0& 0& 0& 0\\
 2& 0& 0& 0& 2& 0\\
 2& 0& 0& 2& 0& 1\\
 0& 0& 0& 0& 1& 2\\
\end{pmatrix} ;
\ \ \
\text{det}(K_{6C}) = -36 , 
\end{align}
\begin{align}
 \label{K6D}
K_{6D} &=
\begin{pmatrix}
 0& 0& 0& 2& 2& 0\\
 0& 0& 1& 0& 0& 0\\
 0& 1& 0& 0& 0& 0\\
 2& 0& 0& 0& 2& 0\\
 2& 0& 0& 2& 0& 0\\
 0& 0& 0& 0& 0& 4\\
\end{pmatrix} ;
\ \ \
\text{det}(K_{6D}) = -64 ,
\end{align}
\begin{align}
 \label{K6E}
K_{6E} &=
\begin{pmatrix}
 0& 0& 0& 0& 2& 0\\
 0& 0& 1& 1& 0& 0\\
 0& 1& 0& 1& 0& 0\\
 0& 1& 1& 0& 0& 0\\
 2& 0& 0& 0& 0& 0\\
 0& 0& 0& 0& 0& 2\\
\end{pmatrix} ;
\ \ \
\text{det}(K_{6E}) = -16 . 
\end{align}
The sixth $K$-matrix that describes the anyon superconductor is given by
\begin{align}
 \label{K6F}
K_{6F} = 2 \del_{IJ},\ \ \ I,J=1,\cdots,6;\ \ \ \ 
\text{det}(K_{6F}) = 64 
.
\end{align}
Such a $K$-matrix quantum Hall state has a higher energy.  Unlike the other
anyon superconductors, the above  anyon superconductor does not break the
permutation symmetry of the six-species of anyons.

\subsection{Anyon superconductivity for three species of anyons}

\def\arraystretch{1.3}
\begin{table*}[tb] 
\caption{Physical properties of the anyon superconductors for
6- and 3-species of anyons.
The table lists 
the number $N_\text{top}$ of topological excitations (with no  $A_\mu$-flux \ie no vorticity),
the minimal $A_\mu$-flux quantum,
chiral central charge $c$ of the fermionic topological order,
the $\mathcal{T}$-matrix,
the $\mathcal{S}$-matrix,
the anyon densities $\nu_I$,
the total energy,
the type of superconductor.
Note that a quasi-particle has $Q_{\vc l}=0$
(\ie has no vorticity).
 }
\label{SFtable} \centering \setlength{\tabcolsep}{3pt}
\begin{tabular}{|c|c|c|c|c|c|c|c|c|}
\hline
$K$-matrix & $N_\text{top}$  & $A_\mu$-flux & $c$  & $\mathcal{T}$-matrix & $\mathcal{S}$-matrix & $\nu_I$ & $E_\text{tot}$ &  type of superconductor\\
\hline
$K_{6A}$ \eqref{K6A} & 1 & $1/2$ & 0 & 1 & 1 & $\frac12,\frac12,\frac12,\frac12,\frac12,\frac12$ & 7.50 & electron-pair $s$-wave \\
\hline
$K_{6A'}$ \eqref{K6Ap} & 1 & $1/2$ & $-2$ & 1 & 1 & $\frac12,\frac12,\frac12,\frac12,\frac12,\frac12$ & 7.50 & electron-pair $g$-wave \\
\hline
$K_{6B}$  \eqref{K6B}& 1 & $1/4$ & 0 & 1 & 1 &$\frac14,1,1,\frac14,\frac14,\frac14$  & 7.60 & 4-electron condensation \\
\hline
$K_{6C}$  \eqref{K6C}& 1 & $1/6$ & 0 & 1 & 1 &$\frac16,1,1,\frac16,\frac13,\frac13$  & 7.63 & 6-electron condensation \\
\hline
$K_{6D}$ \eqref{K6D} & 4 & $1/2$ & 0 & $\mathcal{T}_{\mathbb{Z}_2}$ \eqref{Z2_ST} & $\mathcal{S}_{\mathbb{Z}_2}$ \eqref{Z2_ST} &$\frac14,1,1,\frac14,\frac14,\frac14$  & 7.70 & $\Z_2$-topological order \\
\hline
$K_{6E}$ \eqref{K6E} & 4 & $1/2$ & 0 & $\mathcal{T}_{\text{sem}}\otimes \mathcal{T}_{\text{sem}}$ \eqref{semion_ST} & $\mathcal{S}_{\text{sem}}\otimes \mathcal{S}_{\text{sem}}$ \eqref{semion_ST} &$\frac12,\frac12,\frac12,\frac12,\frac12,\frac12$  & 7.75 & double-semion top. order \\
\hline
$K_{6F}$ \eqref{K6F} & 16 & $1/2$ & 6 & $\mathcal{T}_{\mathbb{Z}_2} \otimes \mathcal{T}_{\mathbb{Z}_2}$ \eqref{Z2_ST} & $\mathcal{S}_{\mathbb{Z}_2} \otimes \mathcal{S}_{\mathbb{Z}_2}$ \eqref{Z2_ST} & $\frac12,\frac12,\frac12,\frac12,\frac12,\frac12$ & 8.00 & double $\mathbb{Z}_2$ top. order\\
\hline
\hline
$K_{3A}$ \eqref{K3A} & 1 & $1/2$ & 1 & 1 & 1 & $1,\frac32,\frac12$ & 2.87 & electron-pair $d$-wave \\
\hline
$K_{3B}$ \eqref{K3B} & 2 & $1/2$ & 1 &  $\mathcal{T}_{\text{sem}}$ \eqref{semion_ST}& $\mathcal{S}_{\text{sem}}$ \eqref{semion_ST} &$\frac32,\frac12,1$  & 3.33 & single-semion top. order \\
\hline
$K_{3C}$ \eqref{K3C} & 7 & $1/2$ & 1 & $\mathcal{T}_7$ \eqref{7_S} & $\mathcal{S}_7$ \eqref{7_S} &$\frac12,1,\frac32$  & 6.34 &  $K= -\begin{pmatrix}
	4 & 3 \\
	3 & 4
\end{pmatrix}$\\
\hline
\end{tabular}
\end{table*}
Now, let us consider the case of three species anyon. We may still view the
$\frac{\pi}{3}$ anyon as a boson attached to $\frac{2\pi}{3}$ flux.  We then
smear the $2\pi \frac13 $ flux per boson into an uniform effective ``magnetic''
field.  This changes the anyon gas to a boson gas of three species of bosons
with a total filling fraction $\tl \nu =3$.  Using $3\times 3$ $K$-matrices, we
cannot find a $K$-matrix quantum Hall state with non-negative $K^{-1} \vc q$ and
$\tl \nu=\vc q^\top K^{-1} \vc q = 3$.

We, instead, view the $\frac{\pi}{3}$ anyon as a boson attached to
$\frac{2\pi}{3}+4\pi\times \text{integer}$ flux.  We then smear the flux per
boson into an uniform effective ``magnetic'' field.  This changes the anyon gas
to a boson gas of three species of bosons with a total filling fraction $\tl \nu
$. If the bosons form a quantum Hall state described by a $3\times 3$
$K$-matrix, such a state, plus its parent electron $\nu=\frac13$ FQH state, is
described by the following total effective theory 
\begin{align}
\label{LtK}
\cL&=
\frac{\tl K_{mn}}{4\pi} \tl a_{m\mu}\partial_\nu \tl a_{n\lambda}\epsilon^{\mu\nu\lambda}
-\frac{\tl q_m}{2\pi} A_\mu\partial_\nu \tl a_{m\lambda}\epsilon^{\mu\nu\lambda}
\nonumber\\
&\ \ \ \
-
\frac{q_{mI}}{2\pi} \tl a_{m\mu} 
\partial_\nu a_{I \lambda } \epsilon^{\mu\nu\lambda}
+
\frac{K_{IJ}}{4\pi}  a_{I\mu}\partial_\nu 
 a_{J\lambda}\epsilon^{\mu\nu\lambda},
\end{align}
where $I,J=1,\cdots,3$, $m,n=1,\cdots,3$, $\frac{1}{2\pi} \partial_\nu 
a_{I\lambda} \epsilon^{\mu\nu\lambda}$ is the current of $I^\text{th}$ anyons.
The $\tl K$ 
and $\tl{\vc q}$
are given by
\begin{align}
 \tl K =
\begin{pmatrix}
 3 & 0 & 0 \\
 0 & 0 & 1 \\
 0 & 1 & 0 \\
\end{pmatrix}
,\ \ \ \
\tl{\vc q}=
\begin{pmatrix}
 1\\
0\\
0\\
\end{pmatrix}
\end{align}
that also describe the $\nu=\frac13$ parent FQH state of electrons.
$q_{mI}$ has a form
\begin{align}
 (q_{mI}) =
\begin{pmatrix}
 1 & 1 & 1\\
 p_1 & p_2 & p_3\\
 p_4 & p_5 & p_6\\
\end{pmatrix}
,
\end{align}
where $p_1,\cdots,p_6$ are integers, describing different ways of attaching
$4\pi$ flux to bosons.  The total effective theory is described by a total
$K$-matrix
\begin{align}
 K_\text{tot} = 
\begin{pmatrix}
 \tl K & -(q_{mI}) \\
 -(q_{mI})^\top & K \\
\end{pmatrix}.
\end{align}

From the equation of motion $ \prt \cL /\prt a_{m0} = \prt \cL /\prt \tl a_{I0} =
0$:
\begin{align}
 \frac{\tl K_{mn}}{2\pi} \partial_i \tl a_{nj}\epsilon^{ij} &=
 \frac{\tl q_m }{2\pi} \partial_i A_j\epsilon^{ij} 
+ \frac{q_{mI}}{2\pi} \partial_i a_{Ij}\epsilon^{ij} 
\nonumber\\
 \frac{q_{mI}}{2\pi} \partial_i \tl a_{mj}\epsilon^{ij} &=
 \frac{K_{IJ}}{2\pi} \partial_i a_{Jj}\epsilon^{ij} 
.
\end{align}
The above two equations implies that
\begin{align}
\frac{\tl K_{mn}-q_{mI}(K^{-1})_{IJ}q_{nJ}}{2\pi}\prt_i\tl a_{nj}\epsilon^{ij} 
 &= \frac{ \tl q_m }{2\pi} \partial_i A_j\epsilon^{ij}, 
\\
 \frac{K_{IJ}-q_{mI} (\tl K^{-1})_{mn} q_{nJ}}{2\pi}\partial_i a_{Jj}\epsilon^{ij} 
 &= \frac{ q_{mI}(\tl K^{-1}\tl{\vc q})_{m} }{2\pi} \partial_i A_j\epsilon^{ij}, 
\nonumber 
\end{align}
which relates the electron density $\frac{1}{2\pi} \partial_i \tl
a_{1j}\epsilon^{ij}$ to the magnetic field $B =  \partial_i A_j\epsilon^{ij}$.
If the above equation can be satisfied by a finite electron density even for
zero  magnetic field $B=0$, then the effective Lagrangian \eq{LtK}
describes an anyon superconductor, \ie an electron superconducting state.  This is
realized by $K$ and $q_{mI}$ such that
\begin{align}
\tl \La_{mn}=  \tl K_{mn} - q_{mI} (K^{-1})_{IJ} q_{nJ}.
\end{align}
has a zero eigenvalue and the corresponding eigenvector $\tl v_m$ has a non-zero
1st component $\tl v_1 \neq 0$, and such that
\begin{align}
\La_{IJ}=   K_{IJ} - q_{mI} (\tl K^{-1})_{mn} q_{nJ}.
\end{align}
has a zero eigenvalue and the  components of the corresponding eigenvector
$f_I$ are all non-zero and have the same sign: $f_I > 0$.  Note that $f_I$ is
proportional to the density of $I^\text{th}$ anyons.  The second requirement
corresponds to all three species of anyons having positive densities.  Note
that the electron density is given by
\begin{align}
n_e= \frac{\partial_i \tl a_{1j}\epsilon^{ij} }{2\pi} &=
 \frac{(\tl K^{-1})_{1m} q_{mI}}{2\pi} \partial_i a_{Ij}\epsilon^{ij} 
=
(\tl K^{-1})_{1m} q_{mI} \rho f_I 
\nonumber 
\end{align}
where $\rho f_I$ is the density of $I^\text{th}$ anyons.  Thus we also require
that
\begin{align}
\rho_e = (\tl K^{-1})_{1m} q_{mI} \rho f_I > 0 .
\end{align}
From the above, we find that
\begin{align}
 \rho
= \frac{n_e}{ (\tl K^{-1})_{1m} q_{mI} f_I} 
\end{align}
Thus the anyon density is given by (in the unit of $n_e$)
\begin{align}
 \nu_I = \frac{f_I}{ (\tl K^{-1})_{1m} q_{mI} f_I} > 0.
\end{align}
This allows us to estimate the energy of anyon superconductor via \eqref{Etot1}.

There are many $K,\ (q_{mI})$ pairs that satisfy the above conditions.  The
first one with lowest energy is given by
\begin{align}
 \label{K3A}
 K_{3A} =
\begin{pmatrix}
 2& 3& 1\\
 3& 2& 2\\
 1& 2& 6\\
\end{pmatrix}, \ \ \
 (q_{mI})_{3A} =
\begin{pmatrix}
 1& 1& 1\\
 1& 1& 1\\
 1& 1& 1\\
\end{pmatrix}
\end{align}
The above choice of $(q_{mI})_{3a}$ corresponds to viewing the $\frac{\pi}{3}$
anyon as a boson attached to $\frac{2\pi}{3}+4\pi$ flux.  We then smear the
flux $2\pi\frac{7}{3}$ per boson into an uniform effective ``magnetic'' field.
This changes the anyon gas to a boson gas of three species of bosons with a
total filling fraction $\tl \nu =\frac{3}{7}$.  Such a boson gas can form a
$\tl \nu=\frac{3}{7}$ quantum Hall state described by the above $K$-matrix.

Other examples include
\begin{align}
 \label{K3B}
 K_{3B} =
\begin{pmatrix}
 2& 4& 2\\
 4& 2& 0\\
 2& 0& 4\\
\end{pmatrix}; \ \ \
 (q_{mI})_{3B} =
\begin{pmatrix}
 1& 1& 1\\
 1& 1& 1\\
 1& 1& 1\\
\end{pmatrix},
\end{align}
and
\begin{align}
 \label{K3C}
 K_{3C} =
\begin{pmatrix}
 0& 6& 0\\
 6& 0& 0\\
 0& 0& 4\\
\end{pmatrix}; \ \ \
 (q_{mI})_{3C} =
\begin{pmatrix}
 1& 1& 1\\
 1& 0& 1\\
 1& 1& 1\\
\end{pmatrix}.
\end{align}

\subsection{Physical properties of anyon superconducting state}

As discussed above, both six-anyon superconductors and three-anyon
superconductors are described by effective theory \eqref{LCS}, with
$K_\text{tot}$ given in the previous two subsections.
Using such an effective theory, we can calculate the
topological order in the anyon superconductors.

The form of the $K$-matrix indicates that the gapped modes belong to an Abelian
fermionic topological order. We extract out the intrinsic bosonic topological
order by factoring out the trivial fermions so
that the $S$ and $T$ matrices satisfy the modular relations $(ST)^3 = e^{i\pi c
/ 4} S^2$ obeyed by the bosonic topological orders. The 3- and 6-species $K$-matrices produce
topological orders (both singly and stacked) characterized by
\begin{align}
\mathcal{S}_{\text{sem}} = \frac{1}{\sqrt{2}} \begin{pmatrix}
	1 & 1 \\
	1 & -1
\end{pmatrix}, \quad
\mathcal{T}_{\text{sem}} = \frac{1}{\sqrt{2}} \begin{pmatrix}
	1 & 0 \\
	0 & i 
\end{pmatrix}
\label{semion_ST}
\end{align}
\begin{align} 
\mathcal{S}_{\mathbb{Z}_2} = \frac{1}{2} \begin{pmatrix}
	1 & 1 & 1 & 1 \\
	1 & 1 & -1 & -1 \\
	1 & -1 & 1 & -1 \\
	1 & -1 & -1 & 1
\end{pmatrix}, \quad
\mathcal{T}_{\mathbb{Z}_2} = \begin{pmatrix}
	1 & 0 & 0 & 0 \\
	0 & 1 & 0 & 0 \\
	0 & 0 & 1 & 0 \\
	0 & 0 & 0 & -1
\end{pmatrix} 
\label{Z2_ST}
\end{align} 
\begin{align}
	\mathcal{S}_7 = \frac{1}{\sqrt{7}} \exp \left(-8ab\pi \ii /7\right), \ \ \ \mathcal{T}_7 = \delta_{ab} \exp \left( 4a^2 \pi \ii /7\right)
	\label{7_S}
\end{align}
The modular relations between $S$ and $T$ matrices then gives the chiral central charge of the bosonic TO, and we have obtained the full modular data of the topological order. Note that the original fermionic TO only provides the chiral central charge to $\text{mod-}1/2$~\cite{LW150704673}, but the decomposed bosonic TO is defined $\text{mod-}8$. The above results are summarized in Table~\ref{SFtable}, along with the bosonic $K$-matrices corresponding to the $S$ and $T$ matrices given. As in the previous case in Table~\ref{SFtablechiral}, we find possible superfluid states with $4e$ and $6e$ condensations, manifested by its vortex quantization.

\

\section{Summary}

Electron gas in 2-dimension with strong Coulomb interaction will form a Wigner
crystal below a critical density. In this paper, we use Laughlin-type wave
functions \eqref{PsibarK} to construct many chiral superconducting states. In
light of the experimental finding,\cite{HJ240815233} we find that, if the
electron has a flat dispersion $\veps \sim k^4$, some of the chiral
superconducting states may have lower energy then Wigner crystal near the
critical density. 

This is because chiral superconductors have larger momentum fluctuations and
larger inter-particle separation compared to the fully spin-valley polarized
Fermi liquid. Thus, the topological chiral superconductors are favored when the
electron band bottom is very flat. In this limit, the Wigner crystal phase may
also be favored. According to our estimate, we find that chiral superconductors
have energies close to that of fully spin-valley polarized Fermi liquid and
Wigner crystal. Therefore, chiral superconductors, if they do appear, are more
likely to appear near the transition between fully spin-valley polarized Fermi
liquid and Wigner crystal.  All these phases are driven by strong repulsive
interaction; this is why the experimentally observed superconducting phase
\cite{HJ240815233} between fully spin-valley polarized Fermi liquid phase and
Wigner crystal phase may be a topological chiral superconductor discussed in
this paper.  

The low energy effective theory of those superconducting states is derived,
which is used to compute the properties of their corresponding superconducting
states.  We find that chiral superconducting states carry non-trivial
topological order and are usually not in the same phase with any BCS superconductors. Namely, they often
have charge-4 or higher condensation and gapless chiral edge modes.

Certainly, a chiral superconductor induced by pairing instability of the quarter
Fermi liquid is also possible, if there is an effective attractive interaction. The
chiral superconductors induced by Coulomb repulsion have an energy scale $0.1 e^2
n_e^{1/2}/\eps$, which is about $2$meV. Compared to this, the observed superconducting state
\cite{HJ240815233} has a transition temperature about $0.3$K.  Such a low
transition temperature may be due to the strong $U(1)$ phase fluctuations. The time
reversal symmetry breaking in correlated electron orbital motion of chiral superconductors 
should persist beyond the
superconducting transition temperature.

Topological chiral superconductors exhibit several characteristics that
distinguish them from conventional BCS superconductors. The discovery of such a
beyond-BCS topological superconductor would be of significant interest. In particular,
when chiral superconductivity emerges in a system with a flat dispersion
relation, such as $\veps = c_4 k^4$, its energy scale from the Coulomb interaction
is large
$0.1 e^2 \bar n_e^{1/2}/\eps 
=0.04 \Big( \frac{e^2}{\eps}  \Big)^{4/3}c_4^{-1/3} $, 
pointing to a potential pathway for achieving high-temperature superconductivity.

\

We would like to thank 
Xiaodong Xu for helpful discussions last year which motivated the anyon-superconducter part of the work. This
work was partially supported by NSF grant DMR-2022428 and by the Simons
Collaboration on Ultra-Quantum Matter, which is a grant from the Simons
Foundation (651446, XGW). LJ acknowledges the support from a Sloan Fellowship. 
AT was supported by NSF Graduate Research Fellowship grant number 2141064.
Some of the numerical calculations were done on subMIT HPC cluster at MIT.

\appendix

\begingroup
\allowdisplaybreaks

\section{Computations of kinetic and interaction energies}
\label{sec:app}

In this section, we calculate the kinetic energy and interaction energy
of the wave function \eqref{PsibarK} of chiral superconductor.
To calculate kinetic energy, we first compute
the equal-time correlation function for a $I_0^\text{th}$ species of particle
\begin{align}
&\ \ \ \ G_{I_0}(
z^{I_0},
z^{I_0 *},
\tl z^{I_0},
\tl z^{I_0 *}
) 
\nonumber\\
&= 
\int \prod_{I,i} 
\dd^2 z_i^I
\Psi^*(\tl z^{I_0}, \{z_i^I)\}
\Psi(z^{I_0}, \{z_i^I\})
\nonumber\\
&=
\int \prod_{I,i} 
\dd^2 z_i^I
\cN \ee^{-\sum_{i,I}\frac{  |z_i^I|^2}{2l_I^2}}
\ee^{-\frac{ |z|^2}{4l_{I_0}^2}}
\ee^{-\frac{ |\tl z|^2}{4l_{I_0}^2}}
\nonumber\\
&\ \ \ \ 
\prod_{i<j,I} |z_i^I - z_j^I|^{2\bar K_{II}}  
\prod_{i,j,I<J} |z_i^I - z_j^J|^{2\bar K_{IJ}}  
\nonumber\\
&\ \ \ \ 
\prod_{i} (z^{I_0} - z_i^{I_0})^{\bar K_{I_0I_0}^+}  
\prod_{i,I} (z^{I_0} - z_i^I)^{\bar K_{I_0I}^+} 
\nonumber\\
&\ \ \ \ 
\prod_{i} (z^{I_0 *} - z_i^{{I_0} *})^{\bar K_{I_0I_0}^-}  
\prod_{i,I} (z^{I_0 *} - z_i^{I *})^{\bar K_{I_0I}^-} 
\nonumber\\
&\ \ \ \ 
\prod_{i} (\tl z^{I_0 *} - z_i^{{I_0} *})^{\bar K_{I_0I_0}^+}  
\prod_{i,I} (\tl z^{I_0 *} - z_i^{I *})^{\bar K_{I_0I}^+} 
\nonumber\\
&\ \ \ \ 
\prod_{i} (\tl z^{I_0 } - z_i^{{I_0}})^{\bar K_{I_0I_0}^-}  
\prod_{i,I} (\tl z^{I_0 } - z_i^{I})^{\bar K_{I_0I}^-} 
\end{align}
where $\cN$ is the normalization coefficient and
\begin{align}
 \bar K_{IJ} = \bar K_{IJ}^+ +\bar K_{IJ}^-
\end{align}
The above integral 
has sign changes, and it is hard to evaluate it via Monte Carlo method.
In the following, we convert the integral into to one that has no sign changes, via holomorphic extension.
Let us consider a related function
\begin{align}
&\ \ \ \ \tl G_{I_0}(
z_1,
z_2^*,
z_3,
z_4^*
) 
\nonumber\\
&=
\int \prod_{I,i} 
\dd^2 z_i^I
\ee^{-\sum_{i,I}\frac{  |z_i^I|^2}{2l_I^2}}
\nonumber\\
&\ \ \ \ 
\prod_{i<j,I} |z_i^I - z_j^I|^{2\bar K_{II}}  
\prod_{i,j,I<J} |z_i^I - z_j^J|^{2\bar K_{IJ}}  
\nonumber\\
&\ \ \ \ 
\prod_{i} (z_1 - z_i^{I_0})^{\bar K_{I_0I_0}^+}  
\prod_{i,I} (z_1 - z_i^I)^{\bar K_{I_0I}^+} 
\nonumber\\
&\ \ \ \ 
\prod_{i} (z_2^* - z_i^{{I_0} *})^{\bar K_{I_0I_0}^-}  
\prod_{i,I} (z_2^* - z_i^{I *})^{\bar K_{I_0I}^-} 
\nonumber\\
&\ \ \ \ 
\prod_{i} ( z_4^* - z_i^{{I_0} *})^{\bar K_{I_0I_0}^+}  
\prod_{i,I} ( z_4^* - z_i^{I *})^{\bar K_{I_0I}^+} 
\nonumber\\
&\ \ \ \ 
\prod_{i} ( z_3 - z_i^{{I_0}})^{\bar K_{I_0I_0}^-}  
\prod_{i,I} ( z_3 - z_i^{I})^{\bar K_{I_0I}^-} 
\end{align}
We note that $\tl G_{I_0}( z_1, z_2^*, z_3, z_4^*) $ is a holomorphic function
of $z_1,z_3$ and an anti-holomorphic function of $z_2,z_4$.  Such a function can
be determined by its values on the subspace $z_1=z_4$ and $z_2=z_3$.
\begin{align}
&\ \ \ \ \tl G_{I_0}(
z_1,
z_2^*,
z_2,
z_1^*
) 
\nonumber\\
&=
\int \prod_{I,i} 
\dd^2 z_i^I
\ee^{-\sum_{i,I}\frac{  |z_i^I|^2}{2l_I^2}}
\nonumber\\
&\ \ \ \ 
\prod_{i<j,I} |z_i^I - z_j^I|^{2\bar K_{II}}  
\prod_{i,j,I<J} |z_i^I - z_j^J|^{2\bar K_{IJ}}  
\nonumber\\
&\ \ \ \ 
\prod_{i} |z_1 - z_i^{I_0}|^{2\bar K_{I_0I_0}^+}  
\prod_{i,I} |z_1 - z_i^I|^{2\bar K_{I_0I}^+} 
\nonumber\\
&\ \ \ \ 
\prod_{i} |z_2 - z_i^{I_0} |^{2\bar K_{I_0I_0}^-}  
\prod_{i,I} |z_2 - z_i^I |^{2\bar K_{I_0I}^-} 
\nonumber\\
&=
\int \prod_{I,i} 
\dd^2 z_i^I
\ee^{-\sum_{i,I}\frac{  |z_i^I|^2}{2l_I^2}}
\nonumber\\
&\ \ \ \ 
\ee^{\sum_{i<j,I} 2\bar K_{II} \log |z_i^I - z_j^I|  }
\ee^{\sum_{i,j,I<J} 2\bar K_{IJ} \log |z_i^I - z_j^J| } 
\nonumber\\
&\ \ \ \ 
\ee^{\sum_{i} 2\bar K_{I_0I_0}^+ \log |z_1 - z_i^{I_0}| }
\ee^{\sum_{i,I} 2\bar K_{I_0I}^+ \log |z_1 - z_i^I| }
\nonumber\\
&\ \ \ \ 
\ee^{\sum_{i} 2\bar K_{I_0I_0}^- \log |z_2 - z_i^{I_0} | }  
\ee^{\sum_{i,I} 2\bar K_{I_0I}^- \log |z_2 - z_i^I | }
\end{align}
Since $(\bar K_{IJ})$ is a positive definite matrix,
it can be written as 
\begin{align}
 2 \bar K_{IJ} =\sum_{a=1}^{\text{dim}(\bar K)} q_I^a q_J^a 
\end{align}
via Cholesky decomposition.
Here we assume the $\bar K$ is invertible.  If it is not, we can shift $\bar
K^+$ $\bar K^-$ by a small positive matrix to make $\bar K$ invertible.  Now we
can rewrite
\begin{align}
- 2\bar K_{IJ} \log |z^I - z^J|
= - q_I^a q_J^a \log |z^I - z^J|,
\end{align}
which can be viewed as the 2-dimensional Coulomb interaction energy between two
charged particles at $z^I$ and $z^J$.  Each charged particle carries
$\text{dim}(\bar K)$ types of charges, labeled by $a$.  The $I$-particle
carries type-$a$ charge $q_I^a$ and the $J$-particle carries type-$a$ charge
$q_J^a$.

Therefore, part of $\tl G_{I_0}( z_1, z_2^*, z_2, z_1^*) $,
\begin{align}
Z &=
\int \prod_{I,i} 
\dd^2 z_i^I
\ee^{-\sum_{i,I}\frac{  |z_i^I|^2}{2l_I^2}}
\nonumber\\
&\ \ \ \ 
\prod_{i<j,I} |z_i^I - z_j^I|^{2\bar K_{II}}  
\prod_{i,j,I<J} |z_i^I - z_j^J|^{2\bar K_{IJ}}  
\end{align}
is the partition function of the above Coulomb gas at temperature $T =1$.  The
term $\frac{ |z_i^I|^2}{2l_I^2}$ represents the potential energy 
of a species-$I$ particle
produced by an uniform
background charge.  We note that a background charge density $\rho =
-\frac1{2\pi}$ will produce a potential $\frac{ |z_i^I|^2}{4}$.  Therefore, the
background charge density $\rho_a$ must satisfy
\begin{align}
 \frac{ |z_i^I|^2}{2l_I^2} = \sum_a -2\pi \rho_a \frac{ |z_i^I|^2}{4} q_I^a \ \ \to \ \
 \rho_a   
= 
-  \sum_I \tl q_I^a \frac{1 }{\pi l_I^2} 
\end{align}
where the matrix $(\tl q_a^I)$ is the inverse of the  matrix $( q^a_I)$:
\begin{align}
 \sum_I q^a_I \tl q_b^I = \del_{ab} .
\end{align}
The relative densities $n_I$ of species-$I$ particles are fixed by filling fractions
of the effective flux via $\sum_J K_{IJ} n_J = 0$.  One way to see this is by requiring
the kinetic energy to scale linearly in particle number in zero background magnetic field.
The total angular momentum at $N^2$ order is given by
\begin{align}
 \frac12 \sum_{IJ} (N_I \bar K_{IJ}^+ N_J - N_I \bar K_{IJ}^- N_J)
\end{align}
If we change $N_I$ by 1, the change of
total angular momentum is given by
\begin{align}
 \sum_{J} ( \bar K_{IJ}^+ N_J -  \bar K_{IJ}^- N_J)
\end{align}
Thus, we require
\begin{align}
 \sum_{J} ( \bar K_{IJ}^+ N_J -  \bar K_{IJ}^- N_J) = 0
\end{align}
in order for the kinetic energy to be finite.  This is also the species ratio of the 
superfluid mode, which is why it persists at zero external magnetic field.

We can now determine the background charge densities $\rho_a$.
The charge neutrality condition of the Coulomb gas gives
\begin{align}
\sum_I n_I q_I^a + \rho_a = 0.
\end{align}
This equivalently determines the inter-particle spacing parameters $l_J$
\begin{align*}
\sum_{I,a} n_I q_I^aq_J^a &=
- \sum_a \rho_a q_J^a = \frac{1}{\pi l_J^2}
\end{align*}
We see that
\begin{align}
\label{spacing}
\sum_J \bar K_{IJ} n_J  = \frac{1}{2\pi l_J^2} .
\end{align}
which is also the condition fixing the size of each species droplet.
The maximal power of $|z|$ is given by $\sum_J \bar K_{IJ} N_J$, and the most probable radius
occupied by this orbital is
\begin{align*}
    R^2 = 2l_I^2 \sum_J \bar K_{IJ} N_J
\end{align*}
reproducing \ref{spacing}.
If $\bar K$ is invertible, we can now express the densities using
\begin{align}
\label{rhoIK}
n_I  = \sum_J  \frac{(\bar K^{-1})_{IJ}}{2\pi l_J^2} .
\end{align}
If $\bar K$ is non-invertible, we cannot readily express $n_I$ in terms of the $l_I^{-1}$.
However, we must remember that $n_I$ are the parameters fixed topologically by $K$, and 
in contrast $l_I$ and $\bar K$ adjust to minimize energy while keeping these densities fixed.
Since $\bar K_{IJ}$ controls the order of zero between species, we expect the value which 
minimizes energy to depend on $n_I$ and $n_J$, so $\bar K_{IJ} = g(n_I,n_J)$.  If $\bar K$ is non-invertible, this means that
one row of $\bar K$ is linearly dependent on the other rows.  We have
\begin{align*}
    g(n_I,n_J) = \sum_K c_K g(n_K,n_J)
\end{align*}
Further assuming $g$ is separable into $g(n_I)g(n_J)$, we can then extract a relation 
$n_I = g^{-1}\left(\sum_K c_K g(n_K)\right)$.  This is to say that a density may usually 
be determined from the others if an inter-particle distance can be determined from the others, 
and therefore a pseudo-inverse of $\bar K$ may be defined.  In many cases, $\bar K$ fails to 
be invertible because all densities are the same, and therefore all entries $\bar K_{IJ}$ are 
the same and all $l_I$ are also the same.  In this case, we can use a matrix proportional to the
identity to relate $n_I$ and $l_J^{-1}$

The term
$-2\bar K_{I_0I}^+ \log |z_1 - z_i^I| $
can also be viewed as a Coulomb energy
between the $z_1$-particle and a species-$I$ particle
if we assume the
$z_1$-particle carries charge $q_+^a$, which satisfies
\begin{align}
 2\bar K_{I_0I}^+ = \sum_a q_+^a q_I^a 
\ \ \to \ \
q_+^a =  \sum_I 2\bar K_{I_0I}^+ \tl q^I_a .
\end{align}
Similarly, the term
$-2\bar K_{I_0I}^- \log |z_2 - z_i^I| $
can be viewed as a Coulomb energy
between $z_2$-particle and a species-$I$ particle
if we assume the
$z_2$-particle carries charge $q_-^a$, which satisfies
\begin{align}
 2\bar K_{I_0I}^- = \sum_a q_-^a q_I^a 
\ \ \to \ \
q_-^a =  \sum_I 2\bar K_{I_0I}^- \tl q^I_a
.
\end{align}
Therefore, the following is the partition function of the Coulomb gas with two
extra charged particles, $z_1$ and $z_2$, present:
\begin{align}
&\ \ \ \ \tl G_{I_0}( z_1, z_2^*, z_2, z_1^*)
\ee^{-\frac{|z_1|^2}{4} \sum_a -2\pi \rho_a q_+^a 
     -\frac{|z_2|^2}{4} \sum_a -2\pi \rho_a q_-^a} 
\nonumber\\
&= 
\int \prod_{I,i} 
\dd^2 z_i^I
\ee^{-\sum_{i,I} \frac{ |z_i^I|^2}{2l_I^2}}
\ee^{ \frac{\pi |z_1|^2}{2} \sum_a  \rho_a q_+^a 
     +\frac{\pi |z_2|^2}{2} \sum_a  \rho_a q_-^a} 
\nonumber\\
&\ \ \ \ 
\ee^{\sum_{i<j,I} 2\bar K_{II} \log |z_i^I - z_j^I|  }
\ee^{\sum_{i,j,I<J} 2\bar K_{IJ} \log |z_i^I - z_j^J| } 
\nonumber\\
&\ \ \ \ 
\ee^{\sum_{i} 2\bar K_{I_0I_0}^+ \log |z_1 - z_i^{I_0}| }
\ee^{\sum_{i,I} 2\bar K_{I_0I}^+ \log |z_1 - z_i^I| }
\nonumber\\
&\ \ \ \ 
\ee^{\sum_{i} 2\bar K_{I_0I_0}^- \log |z_2 - z_i^{I_0} | }  
\ee^{\sum_{i,I} 2\bar K_{I_0I}^- \log |z_2 - z_i^I | } .
\end{align}
the term 
\begin{align}
&\ \ \ \
-\frac{\pi |z_1|^2}{2} \sum_a \rho_a q_+^a 
-\frac{\pi |z_2|^2}{2} \sum_a \rho_a q_-^a
\nonumber\\
&= |z_1|^2 \sum_{I,J}  \frac{1}{2l_I^2} 
(\bar K^{-1})_{IJ}
\bar K_{I_0J}^+
+ |z_2|^2
\sum_{I,J}  \frac{1}{2l_I^2} 
(\bar K^{-1})_{IJ}
\bar K_{I_0J}^-
\nonumber \\
&= |z_1|^2 \sum_{J} \pi n_J 
\bar K_{I_0J}^+
+ |z_2|^2
\sum_{J}  \pi n_J 
\bar K_{I_0J}^-
\end{align}
is  the interaction energy between $z_{1,2}$-particle and the background charge.
In the above we have used
\begin{align}
 \sum_a - \rho_a q_+^a &= \sum_{a,I,J} \tl q_a^I \frac{1}{\pi l_I^2} 2\bar K_{I_0J}^+ \tl q^J_a
= \sum_{I,J}  \frac{1}{\pi l_I^2} 
(\bar K^{-1})_{IJ}
 \bar K_{I_0J}^+
\end{align}
\endgroup
We remark that the  direct interaction energy between $z_{1}$- and $z_2$-particles,
\begin{align}
-\sum_a q_+^a q_-^a\log |z_2 - z_1 | ,
\end{align}
is not included.

If $\bar K_{IJ}$ is not too large, the Coulomb gas is in the plasma phase.  Due
to the perfect screening of the plasma phase, the partition function $\tl
G_{I_0}( z_1, z_2^*, z_2, z_1^*) \ee^{-\frac{|z_1|^2}{4} \sum_a -2\pi \rho_a
q_+^a -\frac{|z_2|^2}{4} \sum_a -2\pi \rho_a q_-^a}$ only depend on the
difference of the positions $z_1-z_2$ of the added charges, if $z_1,z_2$ are in
the plasma droplet of radius $R$.  Therefore
\begin{align*}
 \tl G_{I_0}( z_1, z_2^*, z_2, z_1^*) &= 
g(|z_1-z_2|)
\nonumber\\
&\ \ \ \ \ee^{\frac{|z_1|^2}{4} \sum_a -2\pi \rho_a q_+^a 
    +\frac{|z_2|^2}{4} \sum_a -2\pi \rho_a q_-^a}
\end{align*}
When $|z_2-z_1|$ is small, $\tl G_{I_0}( z_1, z_2^*, z_2, z_1^*)$ has a form
\begin{align*}
 \tl G_{I_0}( z_1, z_2^*, z_2, z_1^*) &= C(1
+ g_2 |z_1-z_2|^2
+ g_4 |z_1-z_2|^4)
\nonumber\\
&\ \ \ \ \ee^{\frac{|z_1|^2}{4} \sum_a -2\pi \rho_a q_+^a 
    +\frac{|z_2|^2}{4} \sum_a -2\pi \rho_a q_-^a} ,
\end{align*}
 where
$C = $ constant. This implies that
\begin{align*}
&\ \ \ \
 \tl G_{I_0}( z_1, z_2^*, z_3, z_4^*)
\\
&= 
C\big(1
+ g_2 (z_4^*-z_2^*) (z_1-z_3)
+ g_4 (z_4^*-z_2^*)^2 (z_1-z_3)^2\big)
\nonumber\\
&\ \ \ \
\ee^{\frac{z_1z_4^*}{4} \sum_a -2\pi \rho_a q_+^a 
    +\frac{z_3z_2^*}{4} \sum_a -2\pi \rho_a q_-^a}  
\nonumber 
\end{align*}
Finally, we find
\begin{align}
&\ \ \ \
G_{I_0}( z, z^*, \tl z, \tl z^*) 
\\
& =
C\big(1
+ g_2 (\tl z^*-z^*) (z-\tl z)
+ g_4 (\tl z^*-z^*)^2 (z-\tl z)^2\big)
\nonumber \\
&\ \ \ \
\ee^{\sum_{IJ} 
\pi n_J
\bar K_{I_0J}^+
z \tl z^* 
-\frac{ |z|^2}{4l_{I_0}^2}}
\ee^{\sum_{IJ}  
\pi n_J
\bar K_{I_0J}^-
\tl z z^* 
-\frac{ |\tl z|^2}{4l_{I_0}^2}}
\nonumber 
\end{align}
We note that $G_{I_0}( z, z^*,  z,  z^*) =\rho_{I_0}(z)$ is the density profile
of a species-$I_0$ particle.  $\rho_{I_0}(z)$ should be a constant $1/\pi R^2$
in a disk of radius $R$, and should become zero outside the disk.  Indeed, we
find that $G_{I_0}( z, z^*,  z,  z^*)$ is independent of $z$, and thus $C =
\frac{1}{\pi R^2}$.

If the species-$I_0$ particle has a kinetic energy operator $-\prt_x^2-\prt_y^2 = -4
\prt_{z^*} \prt_z$, then the average kinetic energy
is given by
\begin{align}
&\ \ \ \ -4\int_{\pi R^2} \dd^2 z\; \prt_{ z^*} \prt_z
G_{I_0}( z, z^*, \tl z, \tl z^*) \mid_{z=\tl z}
\nonumber \\
&= -4\int_{\pi R^2} \frac{\dd^2 z}{\pi R^2}\
|z|^2 \Big(\sum_{IJ} \pi n_J \bar K_{I_0J}^+ 
-\frac{1}{4l_{I_0}^2}\Big)
\nonumber\\
&
\ \ \ \ \ \ \ \ \ \ \ \ \ 
\Big(\sum_{IJ} \pi n_J \bar K_{I_0J}^- 
-\frac{1}{4l_{I_0}^2}\Big)
-\frac{1}{4l_{I_0}^2}
- g_2
\nonumber \\
&=
\frac{1}{l_{I_0}^2} +4 g_2 = \sum_I 2\pi n_I \bar K_{II_0} +4g_2
\end{align}
In the above calculation, we have required $\frac{1}{l_I^2}$ to satisfy
\begin{align}
\label{KplI}
\sum_{IJ} \bar K_{I_0J}^+ \pi n_J =
\frac14 \frac{1}{l_{I_0}^2}
\end{align}
which implies that
\begin{align}
\label{KmlI}
\sum_{IJ} \bar K_{I_0J}^- \pi n_J=
\frac14 \frac{1}{l_{I_0}^2} .
\end{align}
If such a condition is not satisfied, the
average kinetic energy of the single particle will be of order $R^2/l_I^4$,
which approaches $\infty$ as $R\to \infty$.  This is equivalent to the
condition we found earlier based on angular momentum.

If the species-$I_0$ particle has a kinetic energy operator 
$(\prt_x^2+\prt_y^2)^2 =16 
\prt_{z^*}^2 \prt_z^2$, then the average kinetic energy
is given by
\begin{align}
&\ \ \ \ 16\int_{\pi R^2} \dd^2 z\; \prt_{ z^*}^2 \prt_z^2
G_{I_0}( z, z^*, \tl z, \tl z^*) \mid_{z=\tl z}
\nonumber \\
&= 16\int_{\pi R^2} \frac{\dd^2 z}{\pi R^2}\
\frac{1}{8l_{I_0}^4} + 4g_4
=
\frac{2}{l_{I_0}^4}+64 g_4 
\nonumber\\
&= 2(\sum_I 2\pi n_I \bar K_{II_0})^2 + 64 g_4
\end{align}
assuming  $\frac{1}{l_I^2}$ satisfy
\eqref{KplI} and
\eqref{KmlI}.

The conditions \eqref{KplI} and \eqref{KmlI} imply that
\begin{align}
\sum_{J} K_{IJ} n_J = 0,\ \ \
\text{where} \ \ K = \bar K^+ -  \bar K^-. 
\end{align}
So $n_J$ is an eigenvector of $K$ with zero eigenvalue. 
They also imply
\begin{align}
\sum_{J} \bar K_{IJ} n_J = \frac{1}{2\pi l_I^2},
\end{align}
which determines $\frac{1}{2\pi l_I^2}$, that are always all positive as long as $n_I$ are all positive.

Since $\bar K = 2 \bar K^+ - K$, the condition
$\sum_{IJ} \bar K_{IJ}^+ n_J= \frac12 \bar K_{IJ} n_J$ becomes 
\begin{align}
\sum_{IJ} \bar K_{IJ}^+ n_J= \frac12  ( 2 \bar K^+_{IJ} - K_{IJ})n_J
\end{align}
which is valid for any choices of $\bar K_{IJ}^+$, as long as $n_J$ is an
eigenvector of $K$ with zero eigenvalue. 

Therefore, to obtain a wave function for a chiral superconductor, we first
choose an integral symmetric matrix $K$ with odd diagonal elements,
such that it has a single zero eigenvalue with eigenvector $n_I$
satisfying
\begin{align}
n_I = \text{all positive}.
\end{align}
Then we choose a $\bar K_{IJ}^+$ to satisfy
\begin{align}
& \bar K_{IJ}^+ \geq 0, \ \ \ \
 \bar K_{IJ}^- = \bar K_{IJ}^+ - K_{IJ} \geq 0. 
\end{align}
One choice of $\bar K_{IJ}^+$ and $\bar K_{IJ}^-$ is given by
\begin{align}
 \bar K_{IJ}^+ &=
\begin{cases}
 K_{IJ}, & \text{ if } K_{IJ}>0 \\
 0, & \text{ otherwise} \\
\end{cases}
\nonumber\\
 \bar K_{IJ}^- &=
\begin{cases}
- K_{IJ}, & \text{ if } -K_{IJ}>0 \\
 0, & \text{ otherwise} \\
\end{cases}
\end{align}
Such a choice gives rise to a wave function with all ``unnecessary zeros''
removed. The other choice satisfies
\begin{align}
 \bar K_{IJ}^+ + \bar K_{IJ}^- = \bar K_{IJ} = \text{max}(|K_{IJ}|). 
\end{align}
The second choice has lower energies at low densities.
Our results are valid for any choices of $\bar K_{IJ}^+$ and $\bar K_{IJ}^-$.

To summarize, if species-$I$ particle has a kinetic energy operator $
c_2(\prt_x^2+\prt_y^2) +c_4(\prt_x^2+\prt_y^2)^2 $,
the kinetic energy per particle is given by
\begin{align}
E_\text{kin} 
&=  
  \sum_I c_2 \frac{N_I}{N} \left(
  \sum_J 2\pi n_J \bar K_{IJ}+4g_{2, I} 
  \right)
\nonumber\\
&\ \ \ \
+ \sum_I c_4 \frac{N_I}{N} \left(
2(\sum_J 2\pi n_J \bar K_{IJ})^2 +64g_{4, I}
\right)
\nonumber\\
&=   2\pi n_e c_2 \left(
\sum_{IJ}  f_I \bar K_{IJ} f_J +\sum_I f_I \frac{2g_{2, I}}{\pi n_e}
\right)
\nonumber\\
&\ \ \ \
+ (2\pi n_e)^2 c_4 
\left(
\sum_I 2 f_I(\sum_J f_J \bar K_{IJ})^2 + \sum_I f_I \frac{16g_{4, I}}{\pi^2n_e^2}
\right)
\nonumber\\
&=
 2\pi n_e c_2 Z_2
+(2\pi n_e)^2 c_4 Z_4,
\label{KE}
\end{align}
where, 
\begin{align}
\label{Zs}
 Z_2 &\equiv \sum_{IJ}  f_I \bar K_{IJ} f_J + \sum_I f_I \frac{2g_{2, I}}{\pi n_e} 
\\
 Z_4 &\equiv \sum_I 2 f_I(\sum_J f_J \bar K_{IJ})^2 + \sum_I f_I \frac{16g_{4, I}}{\pi^2n_e^2} . 
\nonumber 
\end{align}

Now let us compute the interaction energy
between a species-$I_0$ particle and
a species-$J_0$ particle:
\begin{align}
U_{I_0J_0}
= 
\int \prod_{I,i} 
\dd^2 z_i^I V(z^{I_0} - z^{J_0})
|\Psi(z^{I_0}, z^{J_0}, \{z_i^I\})|^2
\end{align}
Let us introduce the 
density correlation function
\begin{align}
\frac{g_{I_0J_0}(z^{I_0} - z^{J_0})}{(\pi R^2)^2}
= 
\int \prod_{I,i} 
\dd^2 z_i^I 
|\Psi(z^{I_0}, z^{J_0}, \{z_i^I\})|^2
\end{align}
Since $g_{I_0J_0}(z^{I_0} - z^{J_0})$ becomes a constant when $|z^{I_0} -
z^{J_0}|$ is larger than a finite correlation length, and since 
\begin{align}
\int_{\pi R^2} 
\dd^2 z^{I_0}
\dd^2 z^{J_0} \
\frac{g_{I_0J_0}(z^{I_0} - z^{J_0})}{(\pi R^2)^2} =1,
\end{align}
we see that $g_{I_0J_0}(z^{I_0} - z^{J_0}) =1$ when $|z^{I_0} - z^{J_0}|$ is
larger than the correlation length.
We find
\begin{align}
 U_{I_0J_0} =
\int_{\pi R^2} 
\dd^2 z^{I_0}
\dd^2 z^{J_0} \  V(z^{I_0} - z^{J_0})
\frac{g_{I_0J_0}(z^{I_0} - z^{J_0})}{(\pi R^2)^2}
\end{align}
For Coulomb interaction, we also need include background charge:
\begin{align}
 U_{I_0J_0} &=
\int_{\pi R^2} 
\dd^2 z^{I_0}
\dd^2 z^{J_0} \  V(z^{I_0} - z^{J_0})
\frac{g_{I_0J_0}(z^{I_0} - z^{J_0})-1}{(\pi R^2)^2}
\nonumber\\
&=
\int
\dd^2 z \ \frac{e^2}{\eps |z|} 
\frac{g_{I_0J_0}(z)-1}{\pi R^2}
\end{align}
The interaction energy per particle is
\begin{align}
E_\text{int} &= \frac1{2N} \sum_{IJ} U_{IJ} N_I N_J
\nonumber\\
&=
\frac12 \sum_{IJ} f_I f_J n_e\int
\dd^2 z \ \frac{e^2}{\eps |z|} (g_{IJ}(z)-1)
\end{align}
The typical separation between
particles is $n_e^{-1/2}$.
Thus, we rewrite the above as
\begin{align}
 E_\text{int} 
&=
\sum_{IJ} f_I f_J 
\frac{e^2 \sqrt{n_e}}{2\eps}  
\int
\dd^2 z \ \frac{ \sqrt{n_e}}{ |z|} (g_{IJ}(z)-1)
\nonumber\\
&=
\frac{e^2\sqrt{n_e}}{\eps}  
\sum_{IJ} f_I f_J V_{IJ}
=
\frac{e^2\sqrt{n_e}}{\eps} V 
\label{eq:EandV}
\end{align}
where
\begin{align}
 V_{IJ} &\equiv \int
\dd^2 z \ \frac{ \sqrt{n_e}}{2|z|} (g_{IJ}(z)-1) .
\nonumber\\
V &\equiv \sum_{IJ} f_I f_J V_{IJ}.
\label{eq:totenergy}
\end{align}

Let us consider the Coulomb energy for a species-$I$ electron
\begin{align}
E_{\text{int},I} = \frac{e^2\sqrt{n_e}}{\eps}  \sum_J V_{IJ} f_J .
\end{align}
To estimate such a Coulomb
energy, we note that a species-$I_0$ electron behave like a charge $q_{I_0}^a$
particle in the Coulomb gas model of the many-body wave function.  Such a
particle will create a hole of area $A_J$ for the species-${I_0}$ electrons due
to the charge-neutral condition of the plasma phase of the  Coulomb gas.  $A_J$
satisfies
\begin{align}
 \sum_J A_J n_J q_J^a = q_{I_0}^a, 
\end{align}
Since $q_J^a$ is an invertible matrix, We find that, on average,
\begin{align}
 A_I n_I = \del_{II_0}  .
\end{align}
So the species-$I_0$ electron density has a hole of size $\sqrt{1/n_I}$, while
the average density of other electrons is not changed.  This allows us to
estimate
\begin{align}
E_{\text{int},I} = \frac{e^2\sqrt{n_e}}{\eps}  \sum_J V_{IJ} f_J
\approx
-\frac{e^2}{\eps \sqrt{1/n_I}}
\end{align}
or
\begin{align}
V_{IJ} \approx
 - \frac{1}{\sqrt{f_I}} \del_{IJ}  
\end{align}

In fact, even though the
average density of other electrons is not changed, the density of other species electrons should also has a hole of size
$\sim \sqrt{1/n_e}$, if $K_{II_0} \neq 0$ due to the repulsive interaction in the Coulomb gas model.  Including such an
interaction effect, we have an improved estimate 
\begin{align}
V_{IJ} \approx
\begin{cases}
 \frac{1}{\sqrt{f_I}}(a_1 + \frac{a_2 }{\bar K_{II} + a_3}),& I=J\\
 \Theta(\bar K_{IJ}) (a_4 + \frac{a_5 }{\bar K_{IJ} + a_6}), &  I\neq J\\
\end{cases}
\end{align}
where $\Theta(0) = 0$, $\Theta(x > 0) = 1$, and we have included parameters
$a_1,\cdots,a_6$ to fit numerical calculations.  We find
\begin{align}
\label{V_approx}
V_{IJ} =
\begin{cases}
 \frac{1}{\sqrt{f_I}}(-1.830 + \frac{0.408 }{\bar K_{II} + 0.433}),& I=J\\
 \Theta(\bar K_{IJ}) (-1.093 + \frac{0.117 }{\bar K_{IJ} - 0.596}), &  I\neq J\\
\end{cases}
\end{align}
with error $\sim 0.03$.

\def\arraystretch{0.96}
\begin{table}[tb]
    \caption{$V_{IJ}$ for various single- and double-species $\bar K$, alongside their per electron correlation energy (in unit $ \frac{e^2}{l_B}$). }
    \label{tab:corre}
    \centering
    \begin{tabular}{|c|c|c|c|c|c|c|c|}
    \hline
        $\bar K$ & $f_I$ & $E_{int}$ & $\langle K_{IJ} \rangle$ & $V_{11}$ & $V_{22}$ & $V_{12}$ & $V$ \\
        \hline
        $\begin{pmatrix} 1 \end{pmatrix}$ & $(1)$ & -0.617 & 1 & \multicolumn{4}{c|}{-1.545} \\
        \hline
        $\begin{pmatrix} 2 \end{pmatrix}$ & $(1)$ & -0.472 & 2 & \multicolumn{4}{c|}{-1.671} \\
        \hline
        $\begin{pmatrix} 3 \end{pmatrix}$ & $(1)$ & -0.400 & 3 & \multicolumn{4}{c|}{-1.737} \\
        \hline
        $\begin{pmatrix} 4 \end{pmatrix}$ & $(1)$ & -0.351 & 4 & \multicolumn{4}{c|}{-1.758} \\
        \hline
        $\begin{pmatrix} 5 \end{pmatrix}$ & $(1)$ & -0.319 & 5 & \multicolumn{4}{c|}{-1.788} \\
        \hline
        $\begin{pmatrix} 6 \end{pmatrix}$ & $(1)$ & -0.293 & 6 & \multicolumn{4}{c|}{-1.801} \\
        \hline
        $\begin{pmatrix} 1 & 0 \\ 0 & 1 \end{pmatrix}$ & $(\frac{1}{2}, \frac{1}{2})$ & -0.617 & 0.5 & \multicolumn{2}{c|}{-2.185} & 0 & -1.093 \\
        \hline
        $\begin{pmatrix} 2 & 0 \\ 0 & 2 \end{pmatrix}$ & $(\frac{1}{2}, \frac{1}{2})$ & -0.472 & 1 & \multicolumn{2}{c|}{-2.362} & 0 & -1.181 \\
        \hline        
        $\begin{pmatrix} 2 & 1 \\ 1 & 2 \end{pmatrix}$ & $(\frac{1}{2}, \frac{1}{2})$ & -0.519 & 1.5 & \multicolumn{2}{c|}{-2.334} & -0.854 & -1.594 \\
        \hline
        $\begin{pmatrix} 3 & 0 \\ 0 & 3 \end{pmatrix}$ & $(\frac{1}{2}, \frac{1}{2})$ & -0.400 & 1.5 & \multicolumn{2}{c|}{-2.457} & 0 & -1.229 \\
        \hline         
        $\begin{pmatrix} 3 & 1 \\ 1 & 3 \end{pmatrix}$ & $(\frac{1}{2}, \frac{1}{2})$ & -0.446 & 2 & \multicolumn{2}{c|}{-2.405} & -0.755 & -1.580 \\
        \hline
        $\begin{pmatrix} 3 & 2 \\ 2 & 3 \end{pmatrix}$ & $(\frac{1}{2}, \frac{1}{2})$ & -0.426 & 2.5 & \multicolumn{2}{c|}{-2.346} & -1.028 & -1.687 \\
        \hline
        $\begin{pmatrix} 4 & 0 \\ 0 & 4 \end{pmatrix}$ & $(\frac{1}{2}, \frac{1}{2})$ & -0.351 & 2 & \multicolumn{2}{c|}{-2.486} & 0 & -1.243 \\
        \hline         
        $\begin{pmatrix} 4 & 1 \\ 1 & 4 \end{pmatrix}$ & $(\frac{1}{2}, \frac{1}{2})$ & -0.402 & 2.5 & \multicolumn{2}{c|}{-2.474} & -0.715 & -1.594 \\
        \hline
        $\begin{pmatrix} 4 & 2 \\ 2 & 4 \end{pmatrix}$ & $(\frac{1}{2}, \frac{1}{2})$ & -0.385 & 3 & \multicolumn{2}{c|}{-2.412} & -0.926 & -1.669 \\
        \hline
        $\begin{pmatrix} 4 & 3 \\ 3 & 4 \end{pmatrix}$ & $(\frac{1}{2}, \frac{1}{2})$ & -0.359 & 1.67 & \multicolumn{2}{c|}{-2.325} & -1.044 & -1.684 \\
        \hline
        $\begin{pmatrix} 1 & 0 \\ 0 & 2 \end{pmatrix}$ & $(\frac{2}{3}, \frac{1}{3})$ & -0.568 & 0.67 & -1.893  & -2.893 & 0 & -1.163 \\
        \hline
        $\begin{pmatrix} 1 & 0 \\ 0 & 3 \end{pmatrix}$ & $(\frac{3}{4}, \frac{1}{4})$ & -0.562 & 0.75 & -1.784  & -3.475 & 0 & -1.221 \\
        \hline
        $\begin{pmatrix} 1 & 0 \\ 0 & 4 \end{pmatrix}$ & $(\frac{4}{5}, \frac{1}{5})$ & -0.563 & 0.8 & -1.728  & -3.931 & 0 & -1.263 \\
        \hline
        $\begin{pmatrix} 2 & 0 \\ 0 & 3 \end{pmatrix}$ & $(\frac{3}{5}, \frac{2}{5})$ & -0.443 & 1.2 & -2.157  & -2.747 & 0 & -1.216 \\
        \hline
        $\begin{pmatrix} 2 & 1 \\ 1 & 3 \end{pmatrix}$ & $(\frac{2}{3}, \frac{1}{3})$ & -0.496 & 2.55 & -1.997  & -2.963 & -0.876 & -1.606 \\
        \hline
        $\begin{pmatrix} 2 & 0 \\ 0 & 4 \end{pmatrix}$ & $(\frac{2}{3}, \frac{1}{3})$ & -0.431 & 1.33 & -2.046  & -3.045 & 0 & -1.248 \\
        \hline
        $\begin{pmatrix} 2 & 1 \\ 1 & 4 \end{pmatrix}$ & $(\frac{3}{4}, \frac{1}{4})$ & -0.482 & 1.75 & -1.893 & -3.435 & -0.852 & -1.599 \\
        \hline
        $\begin{pmatrix} 3 & 0 \\ 0 & 4 \end{pmatrix}$ & $(\frac{3}{5}, \frac{2}{5})$ & -0.379 & 1.714 & -2.298 & -2.685 & 0 & -1.244 \\
        \hline        
        $\begin{pmatrix} 3 & 1 \\ 1 & 4 \end{pmatrix}$ & $(\frac{3}{5}, \frac{2}{5})$ & -0.433 & 2.2 & -2.215 & -2.771 & -0.767 & -1.609 \\
        \hline
        $\begin{pmatrix} 3 & 2 \\ 2 & 4 \end{pmatrix}$ & $(\frac{2}{3}, \frac{1}{3})$ & -0.419 & 2.67 & -2.038 & -2.977 & -1.074 & -1.714 \\
        \hline
    \end{tabular}
\end{table}

A single MC step constitutes of selecting one electron, and moving its position randomly. The update probability ratio is given by the probability densities of the wavefunction at that configuration, namely
\begin{align}
    &\log \frac{\lvert \psi(z_1, z_2, \cdots, z_i^{new}, \cdots, z_N \rvert^2}{\lvert \psi(z_1, z_2, \cdots, z_i^{old}, \cdots, z_N \rvert^2}   \\
    = & 2 \sum_{j \neq i} K_{IJ} (\lvert z_i^{new} - z_j \rvert - \lvert z_i^{old} - z_j \rvert ) - 
    \frac{\lvert z_i^{new} \rvert^2 - \lvert z_i^{old} \rvert^2} 
    {2l_{IB}^2} 
    .
    \nonumber
\end{align}
When the probability ratio is higher than $1$ (or if the $\log$ value is positive), we accept the change. If the ratio is lower than $1$, then we accept the change with a probability of $\frac{\lvert \psi(z_1, z_2, \cdots, z_i^{new}, \cdots, z_N )\rvert^2}{\lvert \psi(z_1, z_2, \cdots, z_i^{old}, \cdots, z_N )\rvert^2}$.

For the pair distribution function, we only want to extract the bulk properties of the wavefunction. Therefore, we first pick electrons sufficiently inside the liquid, chosen to be within $R/5$ from the origin where $R$ is the radius of the FQH droplet. We then sample the distances between these electrons and all electrons, making sure that for the electron pair in question at least one electron is close to the origin.

We ignore the first $2 \times 10^5$ MC steps for the system to equilibrate, and sample until $3 \times 10^6$ steps for a MC configuration. We have sampled from 128 independent MC configurations.

To verify our potential energy ansatz, we run our Monte Carlo routine for single-species K-matrices from $\bar{K} = 1$ to $\bar{K} = 6$ and for two-species K-matrices $\bar{K} = \begin{pmatrix} a & c \\ c & b \end{pmatrix}$ for $a=1, \cdots, 4$ and $b = 1, \cdots, a$ and $c = 1, \cdots, \text{min}(a, b)$ for 200 electrons.

Looking at the single-species case first, $V_{II}$ can be written as
\begin{align}
 V_{II}(\bar K=m) &= \frac{\pi}{\sqrt{m}}
\int_0^\infty dr \sqrt{\frac{1}{2\pi l_b^2} } \big(g(r)-1 \big).
\end{align}
For $\bar K =1$ the analytical form of the pair distribution function is known:
\begin{align}
g(r) =  1-\ee^{-\frac{r^2}{2l_b^2}}
\end{align}
We find
\begin{align}
 V_{II}(\bar K=1) &= 
- \sqrt{\frac{\pi}{2 l_b^2 }} \int_0^\infty dr \ \ee^{-\frac{r^2}{2l_b^2}}
= -\frac \pi 2 = -1.5707
\end{align}
Our Monte Carlo simulation gave the result $-1.55$, where the small deviation of around 1\% comes comes from the finite-size effects. For the sake of uniformity, we will keep  $N=200$ for all $K$-matrices.
$V_{IJ}$ values and the electron correlation energies for the $K$-matrices considered are given in Table.~\ref{tab:corre}.

\begin{figure}[t]
\hskip -3.0in
\includegraphics[width=0.8\columnwidth]{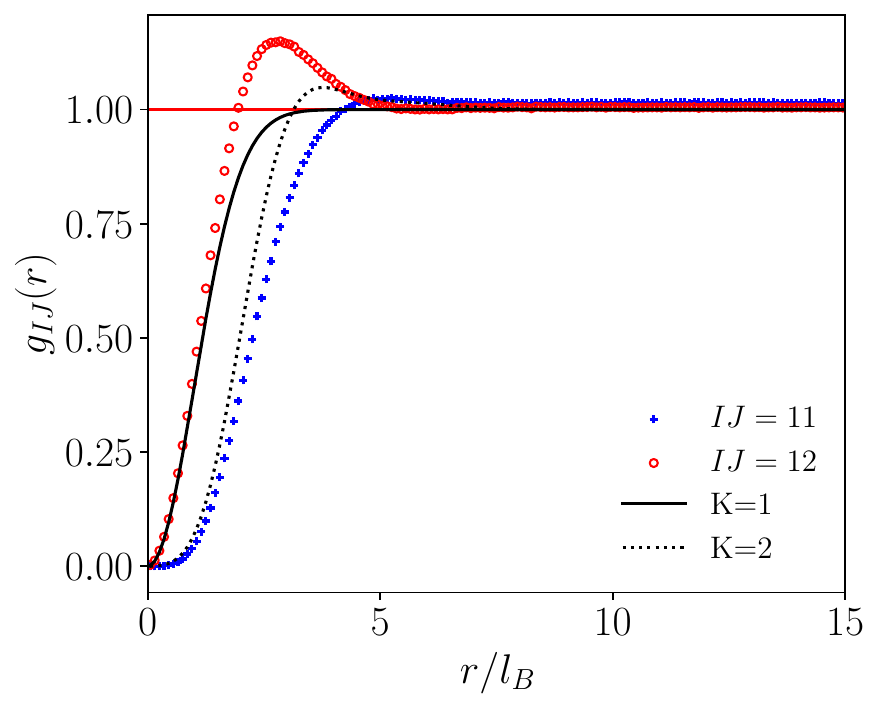}
   \caption{Pair correlation function for intra- and inter-species of $\begin{pmatrix} 2 & 1 \\ 1 & 2 \end{pmatrix}$. The black lines are the pair correlation functions for single-species $K$-matrix $K=1$ (solid) and $K=2$ (dotted).}
    \label{fig:paircorr} 
\end{figure}

Next, we describe our computation of $V_{IJ}$.
For single-species FQH state
\begin{align}
 \Psi(z_i)
= 
\ee^{-\frac{ \sum_i |z_i|^2}{4l_b^2}}
\prod_{i<j} (z_i - z_j)^{\bar K} 
\end{align}
the electron density is given by $n_e = \frac{1}{2\pi l_b^2 \bar K}$. 
Using the Metropolis Monte Carlo method, we can numerically estimate the pair distribution function $g(r)$. For this, we first place the electrons randomly, and follow the Metropolis-Hastings algorithm. 
We extend this to 2-species $K$-matrix. The general 2-species $K$-matrix is
given as $K = \begin{pmatrix} a & c \\ c & b \end{pmatrix}$. For simplicity we
maintain the same $l_B$ values, meaning that the $f_I$ may not be equal to each
other. We maintain the condition that $\nu_I > 0$ for all species. In this
case, the filling fraction is $\nu = \frac{a+b-2c}{ab - c^2}$, giving $n_e =
\frac{\nu}{2\pi l_B^2}$.  For example, Fig.~\ref{fig:paircorr} denotes the pair
correlation function for the $K$-matrix $\begin{pmatrix} 2 & 1 \\ 1 & 2
\end{pmatrix}$, with the black lines being the pair correlation function for
$K=1$ and $K=2$ respectively. We observe that there is higher inter-species
electron pair density and lower intra-species electron pair density.

The reason for this deviation is that as intra-species electrons are more strongly repelled compared to inter-species electrons, compared to the default case the intra-species electrons can spread further from each other due to the lower repulsion with the inter-species electrons, and vice versa. Therefore, intra-species electrons contribute to lower Coulomb energy compared to the default one-species case as they are more further apart whereas the intra-species are more close to each other and contribute more Coulomb energy compared to the single-species case. 
Our numerical results are summarized in Table \ref{tab:corre}.

\bibliography{all, publst }


\end{document}

%% file: defs.tex
\usepackage{graphicx}
\usepackage{graphbox}

\usepackage{dcolumn}
\usepackage{bm}
\usepackage{multirow}

\usepackage[normalem]{ulem}

\usepackage{amsmath}
\usepackage{amsthm}
\usepackage{amstext}
\usepackage{amssymb}
\usepackage{mathrsfs}
\usepackage{amsfonts}
\usepackage{amsbsy} 

\usepackage[all,matrix,cmtip]{xy}

\usepackage{csquotes}
\MakeOuterQuote{"}

\usepackage{color}

\definecolor{red}{rgb}{1,0,0}
\definecolor{blue}{rgb}{0,0,1}
\definecolor{dblue}{rgb}{0,0,0.4}
\definecolor{green}{rgb}{0,1,0}
\definecolor{black}{rgb}{0,0,0}
\definecolor{white}{rgb}{1,1,1}

\definecolor{brn}{rgb}{.8,.4,.0}
\definecolor{redo}{rgb}{1,.5,.0}
\definecolor{ddgrn}{rgb}{0,0.4,0}
\definecolor{dgrn}{rgb}{0,0.55,0}
\definecolor{dbl}{rgb}{0,0,0.5}

\usepackage[bbgreekl]{mathbbol}

\newcommand{\Z}{\mathbb{Z}}

\newcommand{\vc}[1]{\boldsymbol{#1}} 
\newcommand{\tl}[1]{\widetilde{#1}} 
\newcommand{\ii}{\hspace{1pt}\mathrm{i}\hspace{1pt}}
\newcommand{\ee}{\hspace{1pt}\mathrm{e}}
\newcommand{\dd}{\hspace{1pt}\mathrm{d}}
\newcommand{\<}{\langle} 
\renewcommand{\>}{\rangle} 

\newcommand{\Rf}[1]{Ref.~\onlinecite{#1}}

\newcommand{\eq}[1]{(\ref{#1})}

\newcommand{\prt}{\partial}

\newcommand{\up}{\uparrow} 
\newcommand{\down}{\downarrow} 

\newcommand{\ie}{{\it i.e.~}} 
 
\newcommand{\etc}{{\it etc.}}

\newcommand{\bpm}{\begin{pmatrix}}
\newcommand{\epm}{\end{pmatrix}}
\newcommand{\bmm}{\begin{matrix}}
\newcommand{\emm}{\end{matrix}}

\newcommand{\cA}{ {\cal A} }

\newcommand{\cL}{ {\cal L} } 
 
\newcommand{\cN}{ {\cal N} }

\usepackage{euscript}

\newcommand{\al}{\alpha} 
 
\newcommand{\del}{\delta} 
 
\newcommand{\eps}{\epsilon} 
\newcommand{\veps}{\varepsilon} 
\newcommand{\ga}{\gamma} 
 
\newcommand{\ka}{\kappa} 
 
\newcommand{\La}{\Lambda}

\newcommand{\si}{\sigma}